\documentclass[fleqn,useAMS,usenatbib,usegraphicx]{mn2e}
\usepackage{amssymb,amsmath}

\newcommand{\se}[1]{Section~\ref{sec:#1}}

\newcommand{\eq}[1]{equation~(\ref{eq:#1})}

\newcommand{\Eq}[1]{Equation~(\ref{eq:#1})}
\newcommand{\fg}[1]{Fig.~\ref{fig:#1}}

\newcommand{\eqs}[2]{equations\ (\ref{eq:#1}) and (\ref{eq:#2})}

\newcommand{\fgs}[2]{Figs.~\ref{fig:#1} and \ref{fig:#2}}

\newcommand{\Fg}[1]{Figure~\ref{fig:#1}}
\newcommand{\Tb}[1]{\mbox{Table\ \ref{tab:#1}}}
\newcommand{\app}[1]{\mbox{Appendix\ \ref{app:#1}}}
\newcommand\cf{cf.}
\newcommand\eg{e.g.,}
\newcommand\ie{i.e.,}

\newcommand{\pluto}{\textsc{pluto}}

\def\hyph{-\penalty0\hskip0pt\relax}

\newif\ifjournal
\journaltrue

\ifjournal

\newcommand{\?}{}

\newcommand{\gray}[1]{}
\newcommand{\comq}[1]{}
\newcommand{\com}[1]{}
\newcommand{\combox}[1]{}
\newcommand{\comno}[1]{}

\else

\usepackage{color}
\definecolor{gray}{rgb}{0.5,0.5,0.5}
\usepackage{framed}

\newcommand{\?}{\textcolor{red}{$^\textbf{?}$}}

\newcommand{\comq}[1]{$\rightarrow$ \textcolor{red}{#1}}
\newcommand{\com}[1]{[\textcolor{blue}{#1}]}
\newcommand{\gray}[1]{[\textbf{omit}: \textcolor{gray}{#1}]}
\newcommand{\combox}[1]{\begin{framed}\textcolor{blue}{#1}\end{framed}}

\fi

\title{Hydrodynamics of Embedded Planets' First Atmospheres. II. A Rapid Recycling of Atmospheric Gas}
\author[Chris W.~Ormel, Ji-Ming Shi, Rolf Kuiper]{
Chris W.~Ormel$^{1}$\thanks{E-mail:ormel@astro.berkeley.edu}\thanks{Hubble Fellow},
Ji-Ming Shi$^{1,2}$,
Rolf Kuiper$^{3}$\\
$^{1}$Astronomy Department, University of California, Berkeley, CA 94720, USA\\
$^{2}$Department of Astrophysical Sciences, Princeton University, Princeton, NJ\\
$^{3}$Max-Planck-Institut f\"ur Astronomie, K\"onigstuhl 17, D-69117 Heidelberg, Germany
}
\begin{document}
\maketitle
\label{firstpage}
\begin{abstract}
    Following Paper I we investigate the properties of atmospheres that form around small protoplanets embedded in a protoplanetary disc by conducting hydrodynamical simulations. These are now extended to three dimensions, employing a spherical grid centred on the planet. Compression of gas is shown to reduce rotational motions. Contrasting the 2D case, no clear boundary demarcates bound atmospheric gas from disc material;  instead, we find an open system where gas enters the Bondi sphere at high latitudes and leaves through the midplane regions, or, vice versa, when the disc gas rotates sub-Keplerian.  The simulations do not converge to a time-independent solution; instead, the atmosphere is characterized by a time-varying velocity field. Of particular interest is the timescale to replenish the atmosphere by nebular gas, $t_\mathrm{replenish}$. It is shown that the replenishment rate, $M_\mathrm{atm}/t_\mathrm{replenish}$, can be understood in terms of a modified Bondi accretion rate, $\sim$$R_\mathrm{Bondi}^2\rho_\mathrm{gas}v_\mathrm{Bondi}$, where $v_\mathrm{Bondi}$ is set by the Keplerian shear or the magnitude of the sub-Keplerian motion of the gas, whichever is larger. In the inner disk, the atmosphere of embedded protoplanets replenishes on a timescale that is shorter than the Kelvin-Helmholtz contraction (or cooling) timescale. As a result, atmospheric gas can no longer contract and the growth of these atmospheres terminates. Future work must confirm whether these findings continue to apply when the (thermodynamical) idealizations employed in this study are relaxed. But if shown to be broadly applicable, replenishment of atmospheric gas provides a natural explanation for the preponderance of gas-rich but rock-dominant planets like super-Earths and mini-Neptunes.
\end{abstract}


\section{Introduction}
\label{sec:intro}
In this paper, the second of a series, we continue to investigate the properties of the atmospheres that form around low-mass protoplanets as they reside in a circumstellar disc. The low-mass regime that we study entails that the planet's response to the disc falls in the linear regime \citep{LinPapaloizou1993}, but that the planet is massive enough to bind material. The first criterion implies that $R_\mathrm{Bondi}$ is less than the pressure scaleheight of the circumstellar gas $H$; and the latter criterion means that its Bondi radius $R_\mathrm{Bondi}\equiv GM_p/c_s^2$, where $G$ is Newton's constant, $M_p$ the planet mass and $c_s$ the local (isothermal) sound speed of the disc, is larger than the surface radius of the planet, $R_\mathrm{surf}$. Such planets ($R_\mathrm{surf}\le R_\mathrm{Bondi} \le H$) are referred to as `embedded'. Two key questions regarding these atmospheres are: (i)\ does the material stay bound to the planet instead of flowing back to the circumstellar disc; (ii)\ does the flow pattern (the density and velocity field) of the atmosphere attain a steady, time-independent, solution in the frame co-rotating with the planet?

The answer of \citet{OrmelEtal2014} (from here on referred to as Paper I), where we conducted 2D (planar) hydrodynamical simulations using the \pluto\ code \citep{MignoneEtal2007}, to these questions is positive, consistent with earlier findings \citep{KorycanskyPapaloizou1996,Ormel2013}. Paper I also highlighted two features of embedded atmospheres. The first is that the topology of the flow \textit{qualitatively} depends on the value of the so called headwind of the gas -- the amount the gas lags in velocity with respect to Keplerian. This lag arises by virtue of a radial pressure gradient of the circumstellar disc material \citep{AdachiEtal1976,Weidenschilling1997}. If there is no lag (zero headwind) the horseshoe region is centred at the planet's position, which gives rise to large atmospheres. However, in the more realistic case of a pressure gradient, the horseshoe region of a low-mass planet is offset and no longer connects to the atmosphere \citep{Ormel2013}. The bound atmosphere (defined by circulating streamlines) is asymmetrical and much smaller.

Second, we found that the size of the bound atmosphere increases strongly with its mass. That is, as the atmosphere contracts and gas settles towards the centre (\ie\ towards $R_\mathrm{surf}$), the angular momentum of the newly accreted gas causes the atmosphere to spin more rapidly, thereby increasing the size of the circulating -- bound -- region. This is a consequence of vortensity conservation and Kelvin's circulation theorem. We showed that such atmospheres conceivably start to rotate at Keplerian speeds \textit{before} their masses become critical (gas-to-core mass ratio of $\approx$1), which implies a rotational bottleneck for the formation of giant planets.

These findings have been made under the assumption of several idealizations for the flow: primarily, that it is inviscid, isothermal, and 2D (\ie\ planar). In this paper we will relax the latter assumption, which indeed is rather awkward as small planets are fully embedded in the disc. Therefore, the flow is expected to have $z$-component. It is the goal of this paper to conduct 3D hydrodynamical calculations and to contrast it with the 2D-findings of Paper I. 

Three dimensional simulations are, however, computationally more challenging and expensive. In Paper I we saw that a polar grid, with its origin centred on the planet, was advantageous; and we will consider its 3D extension -- spherical grids -- in this work. However, as explained in \se{polar}, grid cells in spherical geometries become very small near the polar angle (the $Z$-axis), and this adds another penalty to the computational efficiency. Another drawback is that 3D flows no longer conserve vortensity. In the 2D simulations vortensity conservation served as a measure to test the fidelity of the hydrodynamical simulations; \ie\ we could quantify the effects of, \eg\ the computational geometry and resolution, simply by following whether the vortensity of the flow deviated from its expected (constant) value. In 3D, however, such a method is not available.

In this paper we will show that the 3D flow is qualitatively different from the 2D case. First of all, the atmosphere carries less rotation, which is attributed to the non-conservation of vortensity. For the parameters adopted in this work, we do not see a transition to a Keplerian-rotating circumplanetary disc. Second, we do not see the emergence of a steady flow -- the gas within the atmosphere is instead prone to fluctuations, small in magnitude but sufficient to alter the integrated quantities (\ie\ streamlines).  Most profoundly, we will show that the atmosphere is no longer bound; that is, atmospheric gas is continually exchanged with gas from the circumplanetary disc. This implies that the replenishment timescale of the atmosphere $t_\mathrm{replenish}$ a factor in the long-term evolution of protoplanetary atmospheres. 

The structure of this paper is as follows: the model, highlighting the differences with the 2D case is presented in \se{model}. Results of the hydrodynamical calculations are presented and analyzed in \se{results}. We discuss these finding and its implications in \se{discuss}. A summary of the key results is given in \se{summary}.

\begin{table*}
  \centering
  \begin{tabular}{lllllll}
  \hline
  \hline
  Name                                   &Domain $(r,\theta,\phi)$                          & Resolution        & Headwind             & Termination     & Expense \\
                                         &                                                  &                   & [$c_s$]   & [$\Omega^{-1}]$      & [CPU days] \\
    (1)                                  & (2)                                              & (3)               & (4)                                   & (5)                             & (6) \\
  \hline
  \texttt{shearSph}                      & [$10^{-3}:0.5]\times\pi/2\times\pi$              & 128x32x64         & $0$     & 10            & 30 \\
  \texttt{shearSph-hiRs}                 & [$10^{-3}:0.5]\times\pi/2\times\pi$              & 256x64x128        & $0$     & 10            & 450 \\
  \texttt{headwSph}                      & [$10^{-3}:0.5]\times\pi/2\times2\pi$             & 128x32x128        & $0.1$   & 10            & 60 \\
  \texttt{headwSph-hiRs}                 & [$10^{-3}:0.5]\times\pi/2\times2\pi$             & 256x64x256        & $0.1$   & 7.4           & 700\\
  \texttt{shearSph-ri7\%}                & [$7\times10^{-4}:0.5]\times\pi/2\times\pi$       & 136x32x64         & $0$     & 10            & 40\\
  \texttt{shearSph-ri7\%-hiRs}           & [$7\times10^{-4}:0.5]\times\pi/2\times\pi$       & 272x64x128        & $0$     & 10            & 700 \\
  \texttt{headwSph-ri7\%}                & [$7\times10^{-4}:0.5]\times\pi/2\times2\pi$      & 136x32x128        & $0.1$   & 10            & 65 \\
  \texttt{headwSph-ri7\%-hiRs}           & [$7\times10^{-4}:0.5]\times\pi/2\times2\pi$      & 272x64x256        & $0.1$   & 6.2          & 1000\\
  \texttt{shearSph-ri5.5\%}               & [$5.5\times10^{-4}:0.5]\times\pi/2\times\pi$    & 210x32x128      & $0$   & 10            & 70 \\
  \hline
  \hline
  \end{tabular}
  \caption{Listing of the runs. The first column gives the name with \texttt{shear} corresponding to a shear-only run, \texttt{headw} a headwind run (see text), \texttt{-hiRs} a run at twice the standard resolution, and \texttt{-riX\%} the value of the inner boundary $r_\mathrm{inn}$ as $X$\% of the Bondi radius (The default is 10\%). The second column gives the domain boundaries as in $[r_\mathrm{inn}:r_\mathrm{out}]\times\theta_\mathrm{out}\times\phi_\mathrm{out}$ with $\theta_\mathrm{in}=\phi_\mathrm{in}=0$. Third column: resolution $N_\mathrm{rad}\times N_\theta \times N_\phi$. When $r_\mathrm{inn}$ is small more grid points are added to keep the grid spacing the same. Column (4): value of the Mach number ${\cal M}_\mathrm{hw}$ of the systematic component of the flow. Column (5): time after which simulations are terminated. Column (6): estimated computational expense
  }
  \label{tab:list}
\end{table*}
\section{Model}
\label{sec:model}
We briefly discuss the setup of the simulation runs, accentuating the differences with the 2D case of Paper I.
\subsection{Units}
Following Paper I, we employ dimensionless units, where lengths are expressed in terms of the scale height of the disc $H$, time in units of the inverse orbital frequency $\Omega^{-1}$, velocities in terms of the (isothermal) sound speed $c_s=\Omega H$ and gas densities in terms of the density in the disc background, $\rho_\mathrm{disc}$. In these units the Bondi radius and the gravitating  mass of the planet $GM_p$ are equal and denoted by $m$: $m = R_\mathrm{Bondi}/H = GM_\mathrm{p}/H^3\Omega^2$. The simulations neglect self-gravity of the gas.

\subsection{Governing equations, domain, and boundary conditions}
Like Paper I we consider a compressible, inviscid and isothermal flow around a gravitating body (planet) on a circular orbit around a star at orbital frequency $\Omega$. We consider the governing equations of this flow -- the Euler and continuity equations -- in a local frame centred on and co-rotating with the planet (see Paper I). The key difference from Paper I is that the gravitational force due to planet $\mathbf{F}_\mathrm{2b}$ has a vertical component. Other forces are the Coriolis force (which does not have a vertical component) and the tidal force. The latter does have a vertical component, ${\cal O}(-z\Omega^2)$ in magnitude. It causes the vertical stratification of the circumstellar disc at scales $\sim$$H$. However, for a low-mass embedded planet this force is negligible. Because the planet mass in this paper is $m=10^{-2}$, we have for simplicity omitted the vertical tidal force.

In Paper I we employed two geometries for the numerical grid: polar $(r,\phi)$ and (uniform) Cartesian $(x,y)$. For the polar geometry we used a logarithmic grid for the radial dimension, which has the advantage that the critical Bondi region is adequately sampled. Both Cartesian and polar geometries led to convergent results but we favored the polar geometry as it was shown to be numerically more efficient than the Cartesian grid. In this paper, we only consider a spherical grid $(r, \theta, \phi)$, where the radial dimension is uniformly sampled in $\log\ r$.

Following Paper I, we again consider \textit{shear-only} runs and \textit{headwind} runs. In the latter, the radial pressure support of the circumstellar disc causes the gas to rotate slower than Keplerian, which adds a systematic and negative component to the unperturbed flow pattern as seen from the planet. That is, the velocity of the unperturbed flow reads $\mathbf{v}_\infty = -({\cal M}_\mathrm{hw} +\frac{3}{2}x)\mathbf{e}_y$ where ${\cal M}_\mathrm{hw}$ is the Mach number of the headwind and $\mathbf{e}_y$ points in the direction of the planet's orbit. For the headwind runs, we consider ${\cal M}_\mathrm{hw}=0.1$.

A summary of simulations is given in \Tb{list}.  Boundary conditions and the domain size for $r$ and $\phi$ are the same as in Paper I. Note that for the shear-only runs only the $y>0$ plane ($0\le\phi\le\pi$) is included in the domain, because of symmetry considerations, whereas for the headwind runs the azimuthal domain consists of the full $2\pi$. In Paper I, the inner radial domain boundary $r_\mathrm{inn}$ was found to be an important parameter as it in essence is a proxy for the mass of the atmosphere. We consider three values of $r_\mathrm{inn}$: $10^{-3}$, $7\times10^{-4}$ and $5.5\times10^{-4}$ or, respectively, 10\%, 7\% and 5.5\% of the Bondi radius. 
 
For the polar ($\theta$) dimension we only follow the upper hemisphere, $z>0$ or $0\le \theta \le \pi/2$, because of the symmetry of the governing equations. The BC at $\theta=\pi/2$ is therefore taken to be reflective: the flow cannot cross the $z$-plane.  More subtle is the BC for $\theta=0$. In \pluto\ we must specify the values of the ghost cells, which lie at $\theta<0$.  Here, we adopt the so-called $\pi$-boundary condition: $Q(r,-\theta,\phi)=Q(r,\theta,\phi+\pi)$ for any quantity $Q$. The $\pi$-boundary simply expresses continuity as one crosses the pole.


Because of the CPU-intensive nature of the simulations, we are unable to conduct a parameter study as diverse as that of Paper I. We decided to fix the planet mass at a dimensionless number of $m=10^{-2}$ (the ratio between the planet's Bondi radius and gas scaleheight of the disc), low enough to ensure that the planet is fully embedded in the disc.  Finally, we conduct these simulations at two grid resolutions: standard and high resolution (\texttt{hiRs}) where in the latter the number of grid points is increased by a factor of 2 in every dimension.

\subsection{The polar angle grid}
\label{sec:polar}
Spherical grids feature two singularities: at $r=0$ and $\theta=0$. The radial $r\rightarrow0$ is not problematic as we only start the grid at the inner radius $r_\mathrm{inn}$ due to the presence of the planet. However, carving out a cone to tackle the polar angle region ($\theta \rightarrow 0$) is unphysical for embedded planets. The problem is that the circumference of the parallels (circles of constant $\theta$) strongly decrease with $\theta$ but that each parallel is subdivided by $N_\phi$ azimuthal grid points. Thus, the minimum grid length in the simulation is of the order of $\Delta l_\mathrm{min} \approx 2\pi r_\mathrm{inn} \sin \theta_1/N_\phi \approx 2\pi r_\mathrm{inn} \theta_1/N_\phi$ where $\theta_1$ is the first (smallest) point in the polar direction and $r_\mathrm{inn}$ the inner radius of the grid. When the $\theta$-points are uniformly distributed this implies a decrease of $\Delta l_\mathrm{min}$ by $1/N_\theta$ compared to the polar grid, which, due to the Courant condition, also implies a subsequent reduction of the timestep.

This is a well-known problem for conducting spherical simulations. A possible solution is to use a cubic spherical grid \citep{KoldobaEtal2002,FragileEtal2009}. As a partial remedy we have adopted a uniform spacing for $\cos \theta$ instead of $\theta$ itself. The uniform spacing in $\cos \theta$ amounts to an equal-area projection and thereby shifts grid points towards lower latitudes. Consequently, the first grid point shifts from $\theta_1\approx \pi/2N_\theta$ for the uniform-$\theta$ spacing to $\theta_1\approx \sqrt{2/N_\theta}$ for the uniform $\cos \theta$. This decreases the computational expense for the high resolution runs ($N_\theta=64$) by a factor 7.

\begin{figure}
  \includegraphics[width=85mm]{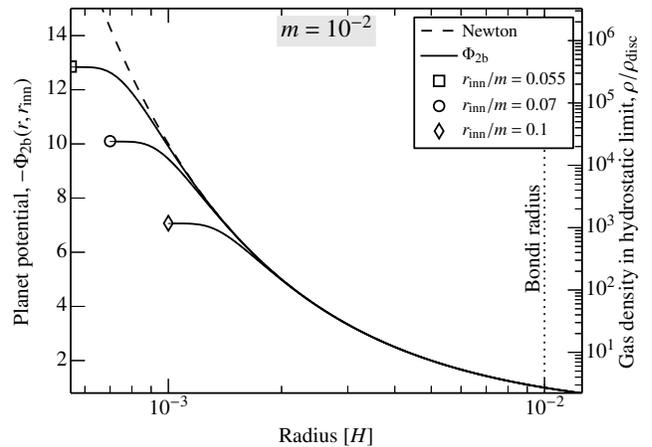}
  \caption{The exponential potential used in this paper, $\Phi_\mathrm{2b}$ (solid curves, \eq{Phi-2b}) for several values of $r_\mathrm{inn}$ and the Newtonian $-m/r$ (dashed). The right axis gives the density in the hydrostatic limit.}
  \label{fig:potentials}
\end{figure}
\begin{figure*}
  \includegraphics[width=180mm]{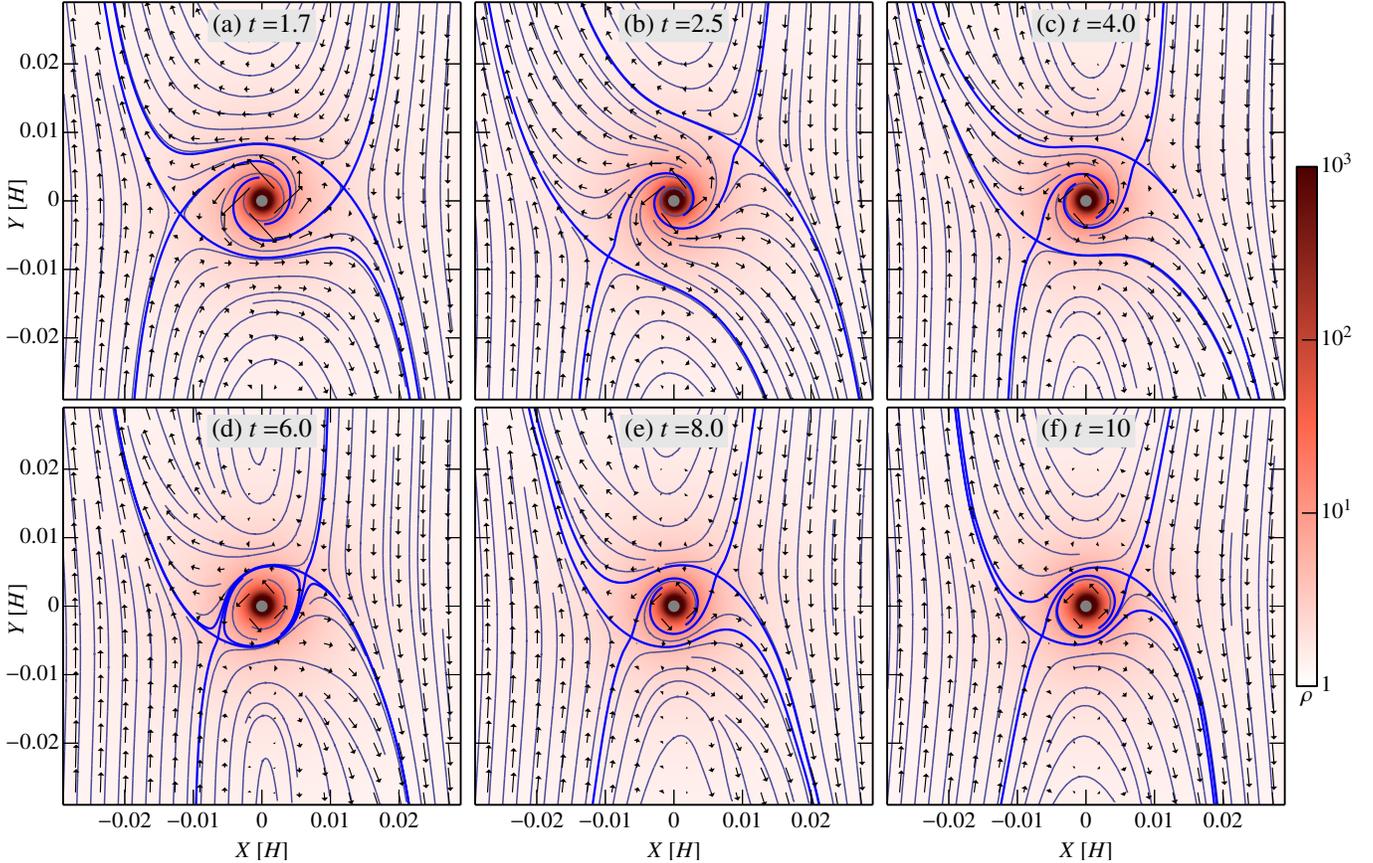}
  \caption{Midplane time sequence of the \texttt{shearSph-ri7\%-hiRs} run, depicting the flow pattern (arrows and streamlines) and the density (shading) in the midplane ($Z=0$). The streamline that crosses the stagnation (X-) point is the critical streamline (thick blue curves). It only exist in the midplane regions where $v_z=0$ by symmetry.}
  \label{fig:stream-evol}
\end{figure*}
\subsection{The two-body force}
In Paper I we adopted a gravitational potential that included a (modest) smoothing length to avoid a singularity in the Cartesian simulations. As we do not consider a Cartesian geometry in this paper, the most natural way is to abolish any softening altogether, \ie\ to use a Newtonian potential.

However, after conducting several test runs we discovered that the Newtonian potential resulted in an unsteady and unphysical flow pattern. We were unable to precisely pinpoint the source of this behavior, but its origin is clearly associated with the large gravitational force at the inner radial boundary. In \app{innerB} we illustrate this phenomenon with a simple 1D application of a damped radial collapse. Clearly, we must modify the boundary condition to include this effect (that is, we found that a simple copying or mirroring of the velocities and densities into the ghost cells to produce reflective or no-slip BCs is not sufficient).

Instead of modifying the BCs, we solve the problem by modifying the gravitational potential such that it yields force-free boundary behavior. A similar procedure has previously been adopted by \citet{AyliffeBate2009i}\com{Other refs?}.  The modified force reads:
\begin{equation}
    F_\mathrm{2b} = -\frac{m}{r^2} \exp\left[ -A \left( \frac{r_\mathrm{inn}}{r} \right)^p \right]
  \label{eq:F2b}
\end{equation}
where $A$ and $p$ are control parameters chosen to be $A=10$ and $p=8$. \Eq{F2b} has the effect that $F_\mathrm{2b}$ very quickly, but continuously, transitions from a pure Newtonian force, $F_\mathrm{2b} = -m/r^2$ for $r\gtrsim 2r_\mathrm{inn}$ to zero for $r\lesssim r_\mathrm{inn}$. 


\begin{figure*}
  \includegraphics[width=160mm]{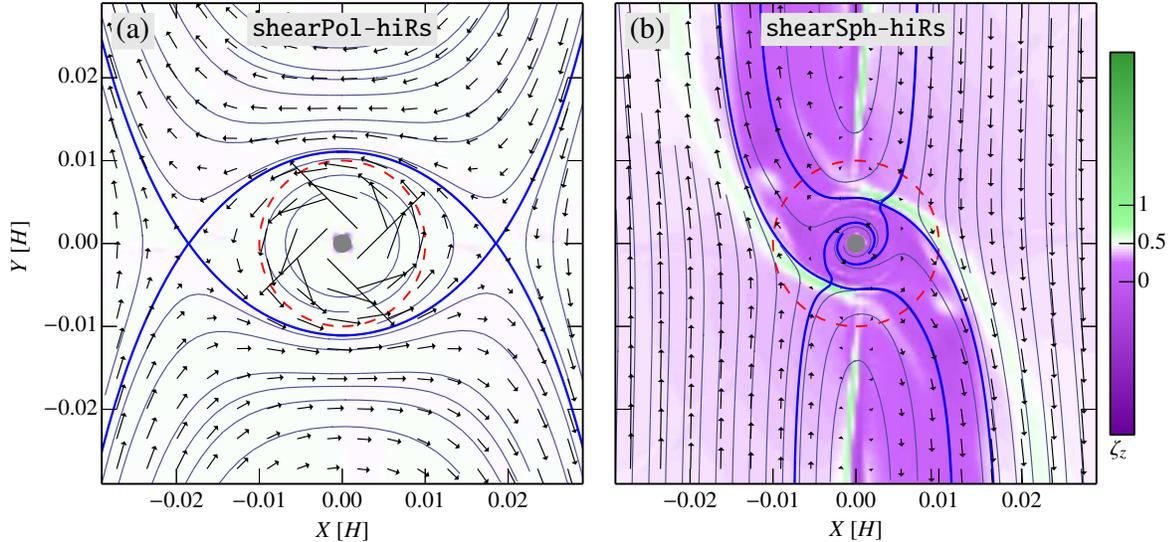}
  \caption{Flow pattern (streamlines and arrows) and vortensity (shading) for the 2D flow of Paper I and the 3D flow in the $Z=0$ plane.  Both runs feature the same mass ($m=10^{-2}$) and inner radius (10\% of $R_\mathrm{Bondi}$). In 3D the vortensity is no longer conserved and generally drops below its unperturbed value of 1/2. As a result the circulation is greatly diminished compared to the 2D case.
  }
  \label{fig:vortensity}
\end{figure*}
The gravitational potential corresponding to \eq{F2b} is
\begin{equation}
  \Phi_\mathrm{2b}
  = \frac{m}{r_0} \frac{\Gamma\left[ 1/p, A (r_0/r)^p \right] -\Gamma(1/p)}{p A^{1/p}}
  \label{eq:Phi-2b}
\end{equation}
where $\Gamma(x)$ is the Gamma function and $\Gamma(a,x)$ the incomplete Gamma function. This function is plotted in \fg{potentials} together with the Newtonian potential $\Phi_\mathrm{Newt}=-m/r$ for the values of $r_\mathrm{inn}$ that we consider in this study: $r_\mathrm{inn}=0.1m$,  $0.07m$, and $0.055m$. For an isothermal EOS the hydrostatic solution of the density is then $\rho = \rho_\mathrm{hs} \equiv \exp[-\Phi_\mathrm{2b}]$, which values are indicated on the right $y$-axis in \fg{potentials}. 

As in Paper I the planet potential is only gradually fed into the simulations. Its temporally dependence is:
\begin{equation}
  \mathbf{F}_\mathrm{2b}(t) 
  = F_\mathrm{2b} \mathbf{e}_r \left( 1 -\exp\left[ -\frac{1}{2} \left( \frac{t}{t_\mathrm{inj}}\right) ^2 \right] \right),
  \label{eq:F2b-time}
\end{equation}
with $t_\mathrm{inj}=0.5$ unless otherwise noted. \Eq{F2b-time} is small initially, but quickly reaches its asymptotic limit for $t\gtrsim2$.

\section{Results}
\label{sec:results}
Below we present and analyze the 3D hydrodynamical simulations. The discussion is organized thematically. The first subsections focus on the differences and similarities with the 2D runs of Paper I: we present the time-evolution in \se{time-evol} and discuss the vortensity (non)conservation in \se{vortensity}. The velocity and density profiles are presented in \se{rhophi} and the statistical properties of the flow in \se{fluctuations}.  The 3D velocity field is presented in \se{3dveloc}. Finally, we analyze the 3D results by presenting a classification scheme for streamlines (\se{topology}) and calculate the replenishment timescale of the atmosphere (\se{replenish}).

\subsection{Emergence to steady state?}
\label{sec:time-evol}
The time-evolution of the density and stream pattern in the midplane ($Z=0$) for the \texttt{shearSph{\hyph}hiRs} run are presented in \fg{stream-evol}. The thick blue curves in \fg{stream-evol} are critical streamlines. These are the streamlines that cross a stagnation point of the flow -- the point where $\mathbf{v}=0$. The existence of these points is not obvious in 3D; however, at the midplane vertical motions vanish by symmetry and the flow much resembles the 2D case. Thus, we can find the `X-point' and draw the `critical streamlines', which also reside in the $Z=0$ plane. However, contrasting the 2D case, these streamline do not demarcate topological different regions, because the critical streamlines do not extend above the midplane regions.

In \fg{stream-evol} the very earliest phases represent the buildup of the hydrostatic atmosphere ($t\lesssim1.5$) are not shown (see Paper I for that). By $t=1.7$ the planet is at 99.7\% of its final value. 
A clear difference from the 2D case is the absence of closed, circulating streamlines near the planet. Instead, gas is flowing out of the inner regions towards the edges of the horseshoe region. This outflow behavior is the rule as seen in the other panels, although it tends to weaken with time.

After $t=2.5$ the outflow is at its peak. Because the density hardly changes from its near-hydrostatic values (see \se{rhophi}) continuity requires that material is replenished from layers $|Z|>0$. We will see in \se{3dveloc} that the material that flows in originates from high latitudes. This is understandable\? as in the shear-only runs the material at $X=0$ co-rotates with the planet. For the gas directly above the planet there is simply only one way to flow: downwards. However, pressure forces will resist the infall. Material thus spirals in at snail pace -- nowhere near free-fall -- with many revolutions before it finally arrives at the midplane.

A key question is whether the midplane flow pattern will reach a steady solution, even if it is not bound. From \fg{stream-evol} this does not seem to be the case: after $t=6$ the outflow streams have closed just to re-open at $t=8$, after which they narrow once again. While for the 2D case, a steady state emerged quickly after the potential has been fully injected, \ie\ around $t=2$, the slow inspiral motions retard the development of a steady flow: clearly, it can only materialize after a replenishment cycle. For this reasons the integration must be carried out on longer timescales, which has the drawback that numerical viscosity affects the result as well as being computationally intensive.


\begin{figure}
    \includegraphics[width=85mm]{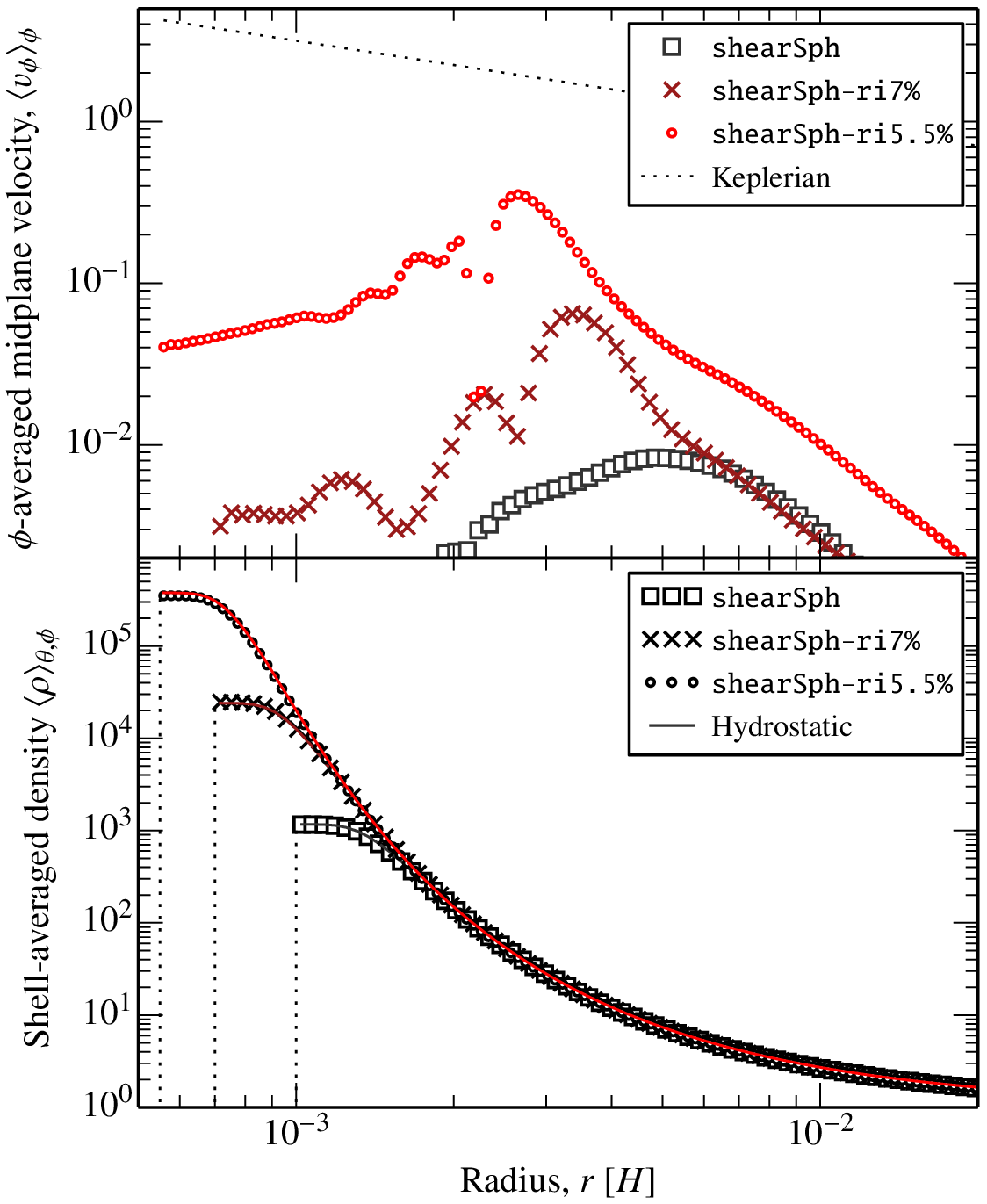}
    \caption{Top: mean azimuthal component of the \textit{midplane} velocity as fucntion of radius for the \texttt{shearSph}, \texttt{shearSph-ri7\%}, and \texttt{shearSph-ri5.5\%} runs at time $t=6$. This shows that velocities increase as the atmosphere contracts and accretes more mass. However, the velocities display a large spread at a fixed $r$ (not shown) and fluctuate strongly in time (see \fg{xcrit}).  Bottom: angularly- (\ie\ shell-) averaged density profile of the atmosphere. The density profiles are constant in time and closely follow the hydrostatic solution (curves). 
    \label{fig:rhophi}
    }
\end{figure}
\subsection{Quenching of the vortensity}
\label{sec:vortensity}
\Fg{vortensity} contrast the flow pattern of the 2D and 3D case. \Fg{vortensity}a shows the flow pattern and the vortensity from one of the runs described in Paper I. The vortensity is defined $\zeta_z = (\omega_z +2\Omega)/\rho$ where $w_z$ is the vertical component of the vorticity, $\boldsymbol{\omega} = \nabla \times \mathbf{v}$. In 2D we see excellent vortensity conservation: $\zeta_z=1/2$ -- the value in the unperturbed state.

\Fg{vortensity}b shows $\zeta_z$ of the 3D simulations in the midplane regions after a time $t=4$. Strikingly, the vortensity is lower, especially near the planet and in the regions where the flow advects to. As a result of the non-conservation of the vortensity, azimuthal velocities are much reduced compared to the 2D case. In the 2D case circulating streamlines demarcate an atmosphere. For the 3D, the analogous topology demarcation would be a closed stream \textit{surface}.  However, 3D flows generally lack a closed stream surface (albeit they may temporarily possess a closed midplane streamline, as is the case in \fg{stream-evol}d). Therefore, the atmosphere in the 3D case is unbound -- as far as boundness is identified with the existence of a closed stream surface.  


We give a heuristic explanation for the vanishing of the vortensity and, hence, the generally lower amounts of rotation compared to the 2D case.  In Paper I we derived the following equation for the vortensity:
\begin{equation}
    \frac{D \boldsymbol{\zeta}}{Dt}
    =  \left( \boldsymbol{\zeta} \cdot \nabla \right) \mathbf{v},
\end{equation}
where $D/Dt$ is the Lagrangian derivative.  In the 2D case the RHS evaluates to 0 because $\boldsymbol{\zeta}$ only has a $z$-component and $\mathbf{v}$ lies in the XY-plane: the vortensity is conserved. In 3D this no longer applies. The vertical component, for example, evolves as
\begin{equation}
    \frac{D \zeta_z}{D t}
    = \zeta_x \frac{\partial v_z}{\partial x} 
        +\zeta_y \frac{\partial v_z}{\partial y} 
        +\zeta_z \frac{\partial v_z}{\partial z}.
\end{equation}
Similar equations apply for $\zeta_x$ and $\zeta_y$: in 3D these are generally non-zero. However, for a streamline initially far from the planet, the unperturbed solution applies and $\zeta_x=\zeta_y=0$ while $\zeta_z=1/2$. 

Ignoring $\zeta_x$ and $\zeta_y$, or assuming that they stay small, we find that $(D\zeta_z/Dt) \propto \zeta_z (\partial_z v_z)$. Now, because of the compression of the gas towards the planet, streamlines converge on each other towards the midplane. This means that the sign of $\partial_z v_z$ is negative and that $D\zeta_z/Dt$ has the opposite sign as $\zeta_z$. Therefore, the vortensity vanishes.

In the fluid dynamical literature, the changes in vortensity are often described in terms of vortex tubes. When these tubes are stretched, the vorticity increases. Vortex stretching plays a central role in the physical explanation of turbulence \citep{TennekesLumley1972}. Here, we encounter the reverse situation: the vortex tubes get compressed and quench the rotation.

\begin{figure*}
  \includegraphics[height=64mm]{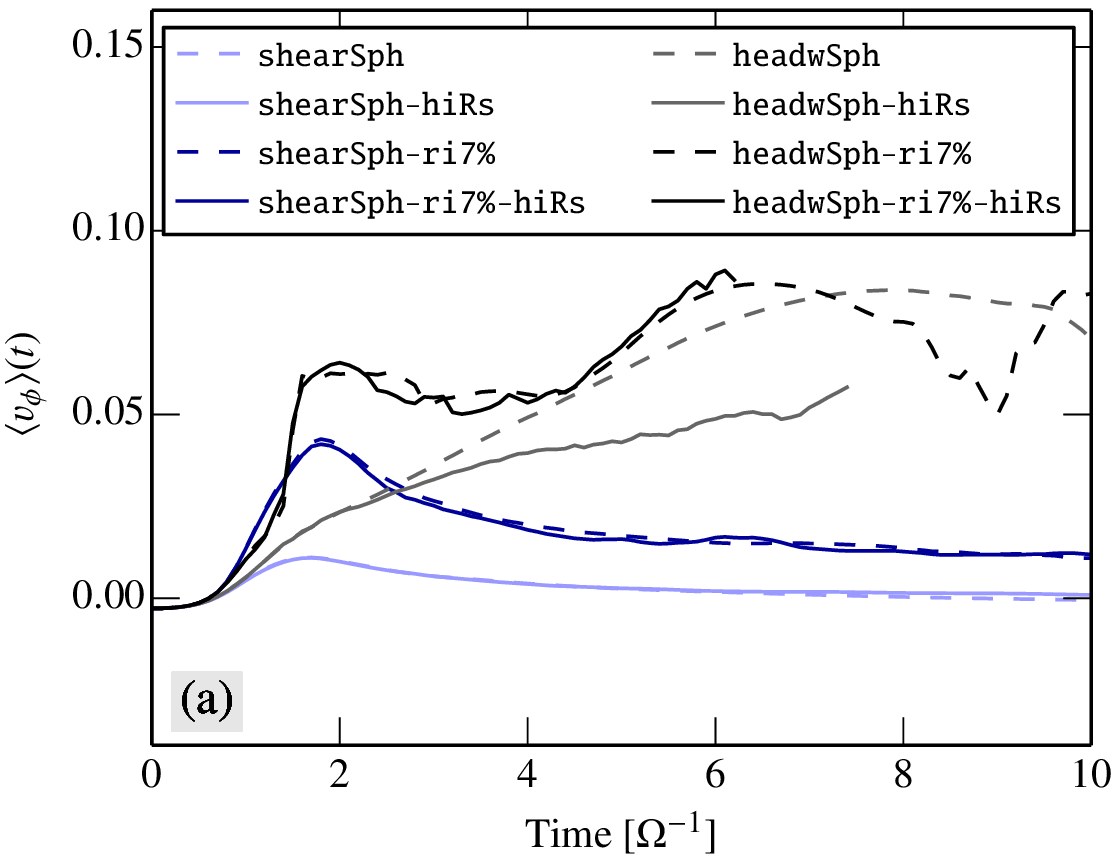}
  \includegraphics[height=64mm]{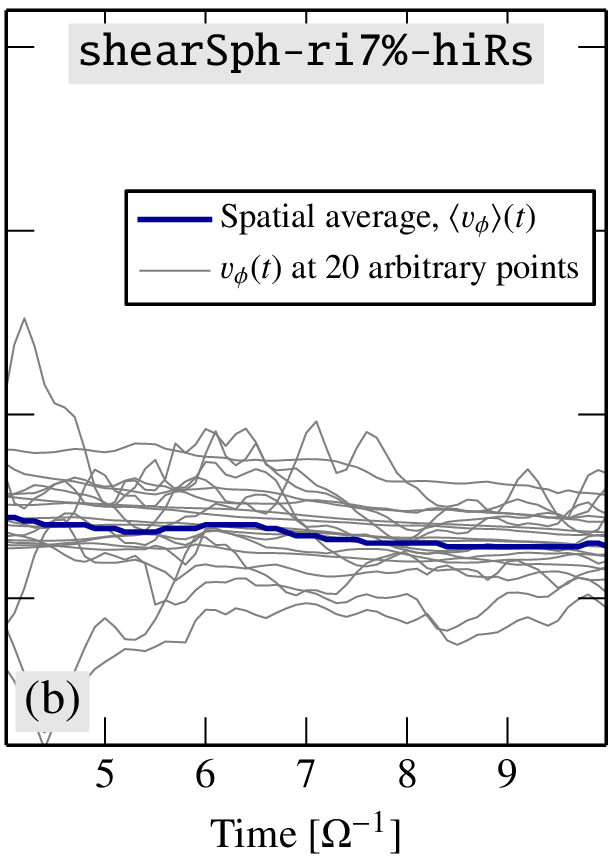}
  \includegraphics[height=64mm]{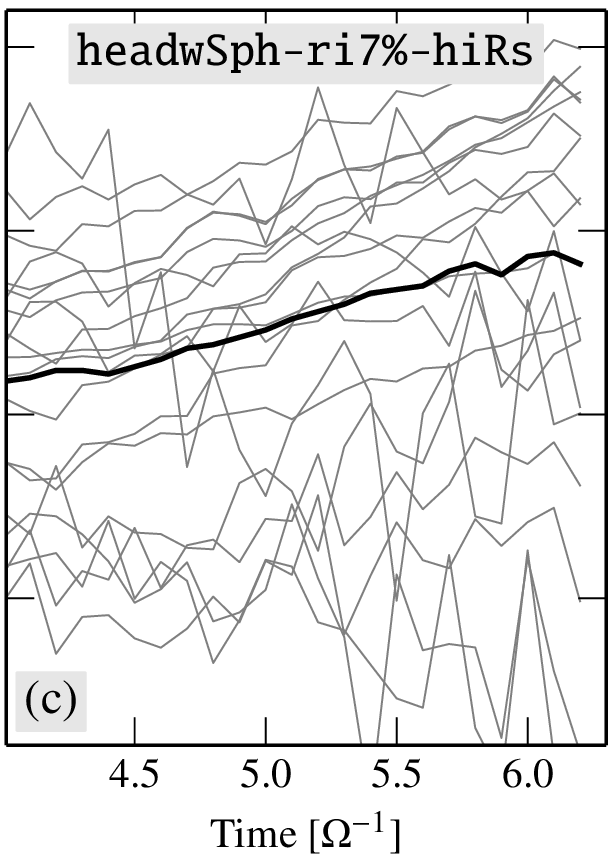}
  \caption{(left:) Spatially-averaged azimuthal velocity $\langle v_\phi\rangle$ in a shell of $0.003 \le r \le 0.006$ for all eight runs. (middle:) Azimuthal velocity at 20 random points within the shell for the \texttt{shearSph-ri7\%-hiRs} run. (right:) Same for the \texttt{headwSph-ri7\%-hiRs} run. These results indicate that, although the spatially-averaged $v_\phi$ evolves smoothly, the velocity at a fixed point can experience significant temporal fluctuations. The $y$-axis is the same for all three panels.}
  \label{fig:xcrit}
  \label{fig:timeQ}
\end{figure*}
\begin{figure*}
  \includegraphics[width=180mm]{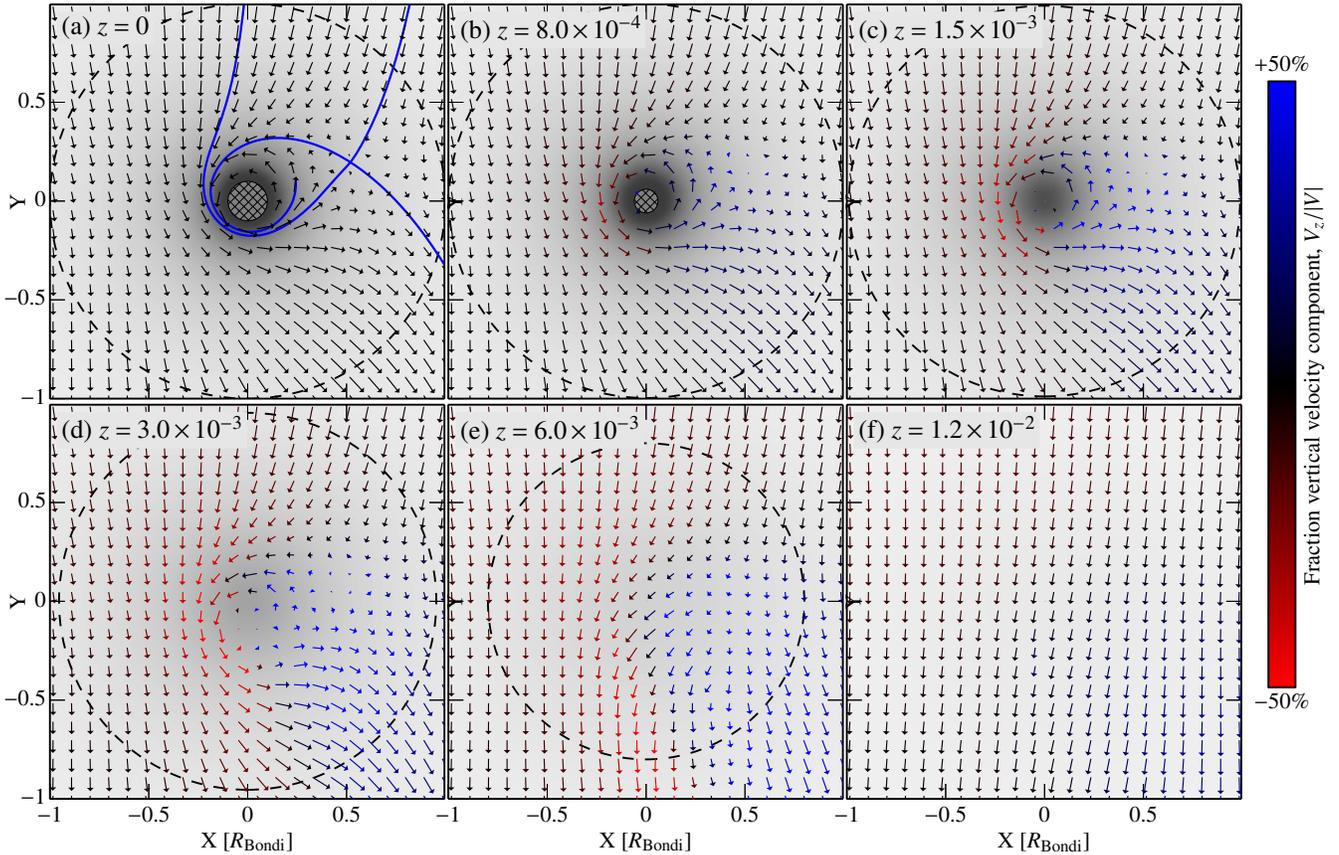}
  \caption{Several slices at fixed $Z$ illustrating the 3D flow structure for the \texttt{headwSph-hiRs} run. The gas density is given in gray in a logarithmic scaling. The intersection with the BS is given by the dashed circle. In panel (a), corresponding to $z=0$, the midplane critical streamline in thick blue shows the 'trap' by which the gas inspirals. By symmetry, there are no vertical velocities at the midplane. At the higher levels (panels (b)--(f)) the vertical motion of the velocity is given by the color of the arrow with blue colors denoting approaching motions (positive $v_z$) and red colors receding motions (negative $v_z$). Spiralling motions extends above the surface of the planet (c) and (d), but at heights above the Bondi radius the flow pattern closely resembles its unperturbed state (f).}
  \label{fig:headw-Zfig}
\end{figure*}
\subsection{Density and rotation profiles}
\label{sec:rhophi}
\Fg{rhophi} shows the density and velocity profiles for three runs that differ only by the value of the inner radius $r_\mathrm{inn}$.  By virtue of the isothermal EOS a decreasing $r_\mathrm{inn}$ implies a strong increase of the atmosphere and mass within  a certain reference radius. As a result, the atmosphere spins up. In the 2D calculations of Paper I we saw that by $r_\mathrm{inn}=0.05m$ the inner regions were tending towards Keplerian rotation and their density profiles had deviated from hydrostatic. However, in the 3D case the tendency towards rotation is suppressed. \Fg{rhophi}a shows that Keplerian rotation is nowhere obtained and \fg{rhophi}b shows that the density follows the hydrostatic solution (solid curves; note that these curves bend over at small $r_\mathrm{inn}$ because of the softening of the potential for $r\rightarrow r_\mathrm{inn}$). Remark that the velocities are time-dependent -- in the shear runs they tend to decrease with time as seen in \fg{stream-evol} -- but gas density is not.

Nevertheless, \fg{rhophi} also shows that qualitatively the trend identified in Paper I -- more massive atmospheres rotate faster -- still holds: the \texttt{shearSph-ri5.5\%} run has a more massive atmosphere and stronger rotational motions than the \texttt{shearSph} curve. Conceivably, these atmospheres may become rotationally dominated when $r_\mathrm{inn}$ decreases further, or when the mass of the planet increases. In Paper I, we identified this point as a centrifugal barrier for the growth of the atmosphere, which, when certain conditions are fulfilled, acts to prevent the atmosphere from reaching a mass similar to that of the core (a.k.a.~the critical core mass).  Conducting simulations at even lower $r_\mathrm{inn}$ is beyond the scope of this work. Computationally, they are very expensive: resolving the steep density gradients imply a very fine radial grid and the large velocities render the explicit hydrodynamical timestep short. In addition, as noted, the flow must be followed for longer timescales.

\subsection{Fluctuations}
\label{sec:fluctuations}
We return to the question whether the 3D flow can be characterized as steady.  In \fg{timeQ}a we plot the spatially-averaged azimuthal velocity of all grid points that lie at a distance between 30\% and 60\% of the Bondi radius. This shows whether the bulk motion of the gas is prograde or retrograde. Initially, at $t=0$ it is retrograde because of the Keplerian shear. However, as gas falls in, towards $t\approx2$, the sense of rotation becomes clearly prograde. Also, one notes that the low $r_\mathrm{inn}$ runs rotate faster, for reasons explained above. However, for the shear runs the rotation then diminishes as the atmosphere is being replenished by material around the polar regions, which is low in angular momentum. As a result, the velocities decrease. For the headwind runs there is no inflow from the polar regions and $v_\phi$ stays large. But in contrast to the shear-only runs it does not flatten out. There may exist an oscillatory pattern, alternating between periods of (relatively) high bulk motion and low bulk motion. Velocitiess are always subsonic.

\Fg{timeQ}a shows agreement between the low and high resolution runs, with the exception being the \texttt{headwSph-hiRs} run, which diverge after $t=3$. Unfortunately, it is beyond the scope of this work to conduct higher resolution simulations.

From the shear-only runs of \fg{timeQ}a one might expect that they have achieved time-independence. However, this is misleading as \fg{timeQ} plots spatially-averaged quantities. In fact, the flow is characterized by significant spatial and temporal fluctuations in $v_\phi$. This is shown in \fg{timeQ}b and c, which plots $v_\phi(t)$ of 20 randomly selected points within the shell. Although some points show a trend consistent with the spatial average, many other points fluctuate strongly. These fluctuations are apparently uncorrelated as they do not affect the spatial average. They are the key reason why the flow cannot be described as steady.

We can quantify  the magnitude of the fluctuations by computing the standard deviations of a time series at a fixed grid point $i$ in the shell, \ie\ $\sigma_i \equiv \mathrm{std}(v_{\phi,i} -\langle v_\phi\rangle)$. Averaging the $\{\sigma_i\}$ we find that the typical fluctuation is at the 10\% level of the bulk motion $\langle v_\phi \rangle$. The velocity field is thus moderately unsteady; and this will have significant implications for the flow pattern. On the other hand, the velocity fluctuations are never strong enough to perturb the density. A similar analysis conducted for $\rho$ shows that the gas density fluctuations are at the 0.1\% level or less.

\subsection{The 3D velocity field}
\label{sec:3dveloc}
The 2D midplane flow pattern, as we have surveyed in \fgs{stream-evol}{vortensity} is merely a projection of the full 3D flow structure. \Fg{headw-Zfig} presents several cuts at fixed heights for the \texttt{headw-hiRs} run. Except at $z=0$, the velocities have a vertical component, which we indicate by a red-to-blue color shading: red corresponding to receding motions, while blue represents a blueshift of the velocity. 

In panel (a) the critical streamline is indicated. It shows, intriguingly, an open configuration: material flows into a rotating region close to the planet surface. This is the reverse of the shear-only runs where the midplane regions saw outflow. This (inflow) configuration of the critical streamline is typical. Although, as we saw in \fg{xcrit}, the position of the X-point and thereby of the critical streamline itself wobbles in time, the open configuration is characteristic of the \texttt{headwind} simulations.

Going to larger heights, we see that the flow structure becomes three dimensional. Very crudely, one notes that the vertical motions in the $+Y$ planes are towards the planet while those at $-Y$ are away from the planet, but also that the right-half planes (at positive $X$) are `bluer' than the left-half. The explanation for the first trend is that upstream material compresses when it approaches it planet. The second trend is a manifestation of the Coriolis force, which swirls material from $-X$ to $+X$, especially in the midplane regions. Material at larger $Z$ then decreases in $Z$ to compensate this removal.

This circulation seen at $z=0$ is repeated, in its projected sense, at heights $\lesssim$$3\times10^{-3}$. It is, admittedly, very arbitrary to define `circulatory' motions based on a projection alone; as we will see, streamline trajectories are very much 3D. However, for circulation to be present the azimuthal velocity ($v_y$) must be positive somewhere in the plane, which it is in panels (a)--(d) but no longer in panel (e). Here, at $z=6\times10^{-3}$, vertical motions are still obvious, but all velocities have a negative $Y$-component, aligned with the headwind. Thus, we can argue that this height -- still well within the Bondi radius -- represents the transition from where the flow is primarily influenced by the planet to where it is influenced by the headwind. By $z=1.2\times10^{-2}$, just out of the Bondi sphere, the planet barely influences the flow pattern.

\begin{figure}
  \includegraphics[width=85mm]{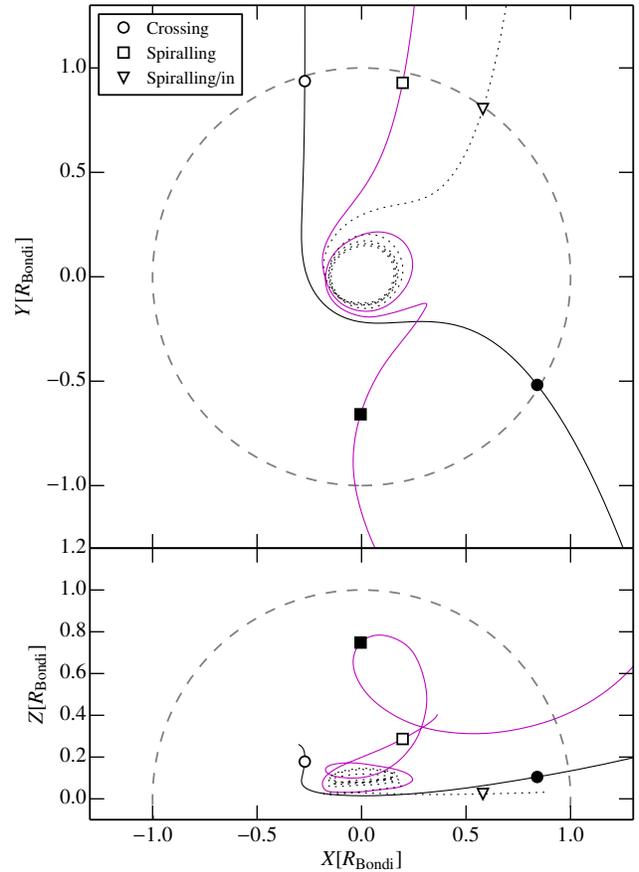}
  \caption{Examples of the two classes of Bondi-sphere (BS) penetrating streamlines for the \texttt{headwSph-hiRs} run. The top panel gives the projection in the XY-plane and the bottom panel in the XZ-plane. Open symbols give the point where the streams enter the BS; closed symbols where they exit. Spiralling streamlines (square and triangle) complete at least one revolution around the $z$-axis. The numerical integration of the stream labeled `Spiralling/in' is terminated after 5 revolutions.
  }
  \label{fig:streams-headw}
\end{figure}
\subsection{The streamline topology}
\label{sec:topology}
We classify streamlines that enter the Bondi sphere (BS) in two categories:
\begin{enumerate}
    \item BS-\textit{spiralling} trajectories. These are streams that enter or leave the Bondi sphere while completing at least one full revolution around the $z$-axis.
    \item BS-\textit{crossing} trajectories. These are streamlines that enter or leave the Bondi sphere without conducting a full revolution.    
\end{enumerate}

Examples of crossing and spiralling streams are shown in \fg{streams-headw} for the \texttt{headwSph-hiRs} run and in \fg{streams-shear} for the \texttt{shearSph-hiRs} run. These figures show the projection on the $XY$ and $XZ$ planes. In these figures the symbols denote the position where the streams enter or leave the BS. Crossing streams trajectories can be strongly influenced by the planet -- like the horseshoe streams in \fg{streams-shear} and occasionally show significant vertical motion -- but they do not complete a full revolution around the origin. Spiralling streams, on the other hand, seem to be influenced even more profoundly by the planet: they tend to come closer to the planet and start to orbit it,  albeit not on closed -- bound -- streamlines.

\begin{figure}
  \includegraphics[width=85mm]{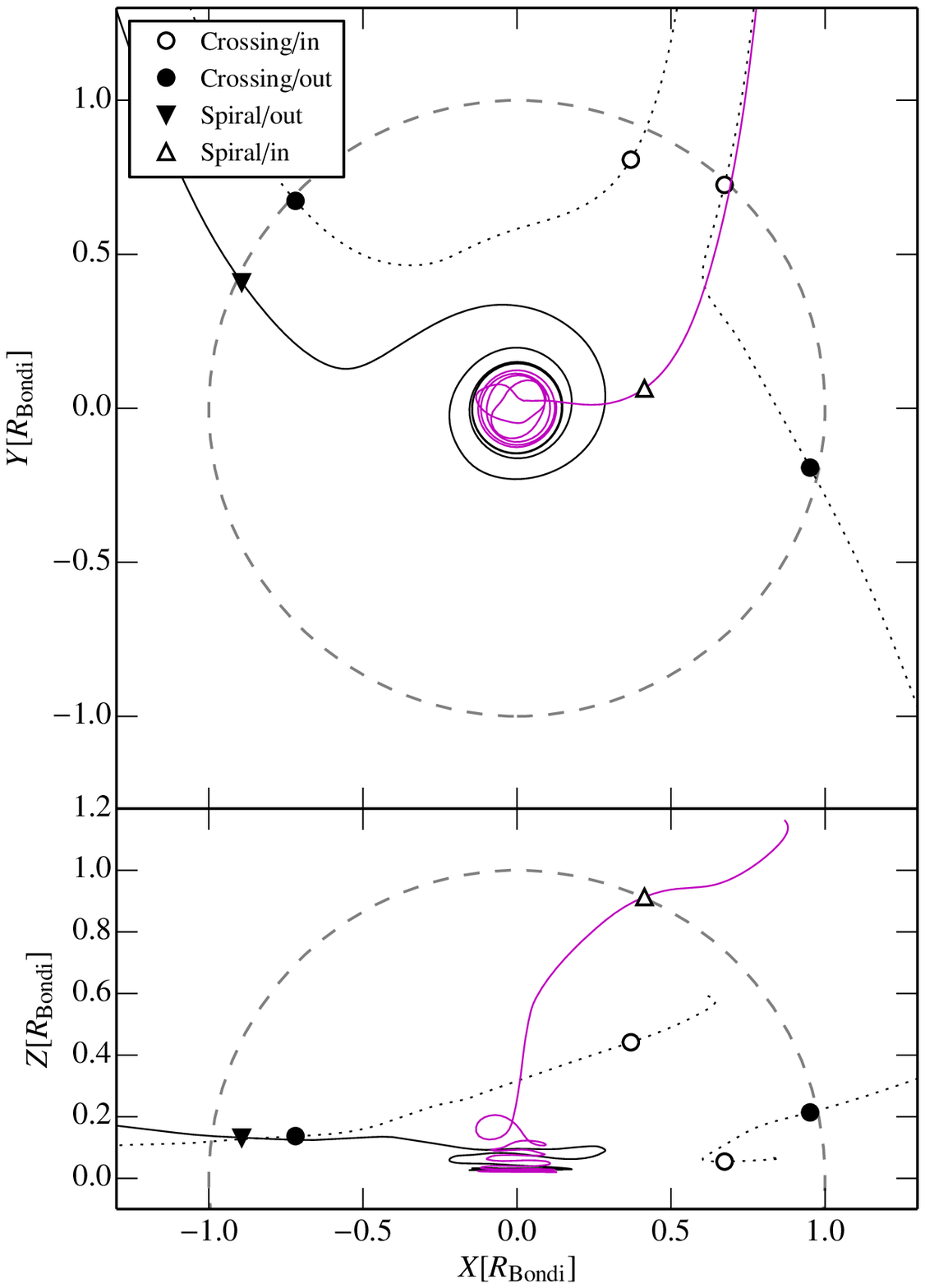}
  \caption{Same as \fg{streams-headw} but for the \texttt{shearSph-hiRs} run. The `Spiral/out' stream is followed stream-upward. The horseshoe orbit is classified as a BS-crossing stream. }
  \label{fig:streams-shear}
\end{figure}
Numerically, we terminate the integration of spiralling streams when the number of revolutions (in either direction) has hit 5. A reason for discontinuing the integration at a certain point is that it may have entered an infinite loop because the flow pattern is not time-steady or because of numerical integration errors. Consequently, the integration terminates within the BS. For this reason we have also conducted upstream integrations. Physically, these streams originate from the central regions and flow out of the BS. But computationally, it is expedient to start from a point at/beyond the BS and integrate backwards to the central regions. In \fg{streams-shear} we have labeled the forward and backward integrated streams as `Spiral/in' and `Spiral/out' respectively.




\begin{figure}
  \includegraphics[height=52mm]{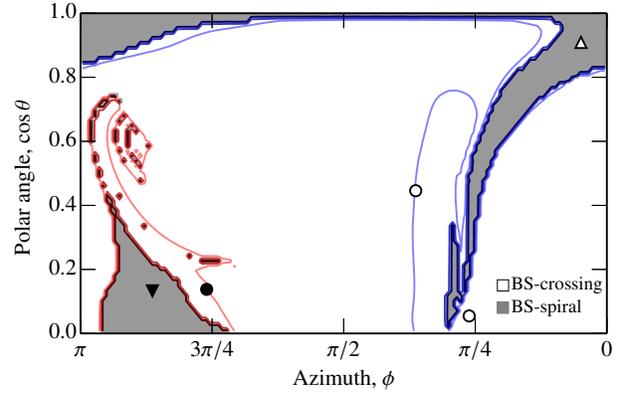}
  \caption{Topology map of the \texttt{shearSph-hiRs} run at time $t=4$\?. The fate of every $N_\theta \times N_\phi$ streamline on the Bondi sphere (BS) is categorized as crossing (empty) or spiralling (grey). In addition, contours provide the minimum distance that each streamline trajectory attains: $r_\mathrm{min}=2\times 10^{-3}$, $4\times10^{-3}$ and $6\times10^{-3}$. Blue contours denote downstream integrations (for streams that enter the BS) while red contours give the $r_\mathrm{min}$ of the streams that emerge from the BS.  Symbols give the same position as in \fg{streams-shear}. Note that the azimuthal range extends only to $\theta=\pi$ from symmetry considerations. Spiralling trajectories originate mostly from high latitudes, penetrate deeply in the atmosphere, to exit the BS at mid-to-low latitudes.
  }
  \label{fig:topol-shear}
\end{figure}

From \fgs{streams-headw}{streams-shear} it is clear that spiralling streams penetrate the atmosphere most deeply.  More systematically, we have classified all streams corresponding to each of the $N_\theta \times N_\phi$ grid points on the BS. Trajectories are always followed into the BS, which means that approximately half of the integrations are integrated in the upstream direction. Like with the streams shown in \fgs{streams-headw}{streams-shear} we follow each trajectory until it exits the BS or until the number of revolutions exceeds 5 (whereafter we stop the integration). Each of these integration are classified as spiralling or crossing. In addition, we record the minimum distance to the planet $r_\mathrm{min}$ that each stream attains.

The result of this classification is presented in \fg{topol-shear} for the \texttt{shearSph-hiRs} run. This shows the fate or origin (for backwards integrations) of the streams on an equal-area (cylindrical) projection of the BS. BS-crossing streams are colored white while BS-spiralling streams are gray. Furthermore, the minimum distance $r_\mathrm{min}$ is indicated by contour lines of $r_\mathrm{min}=20\%$, 40\%, and 60\% of the Bondi radius. The two different color schemes -- blue and red -- indicate that the streams either leave or enter the BS. For \fg{topol-shear} only azimuths of $0\le\phi\le\pi$ are given as the pattern repeats for $\pi\le\phi\le2\pi$. 



\begin{figure*}
  \includegraphics[height=52mm]{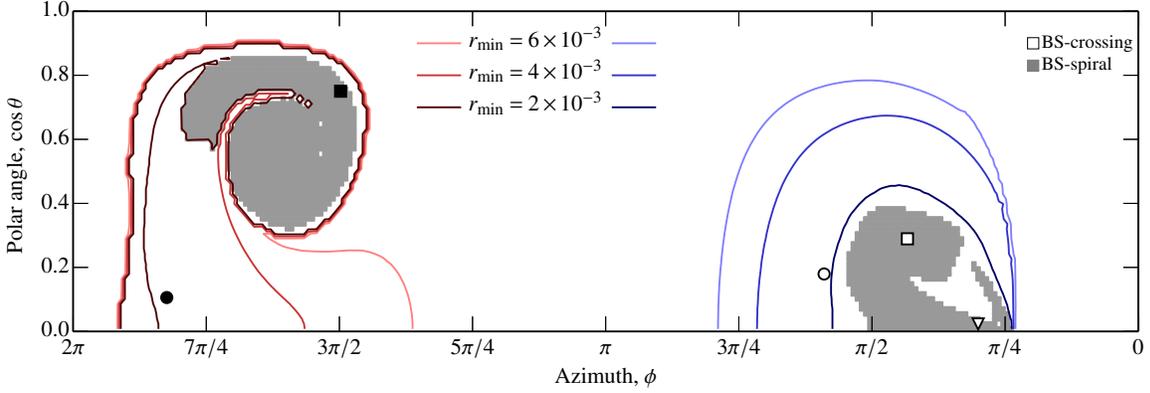}
  \caption{Topology map of the \texttt{headwSph-hiRs} run at time $t=4$. See \fg{topol-shear} for a description. Spiralling trajectories reach closest to the planet core. They preferentially enter the Bondi sphere at low latitudes and leave at mid-to-high latitudes.}
  \label{fig:topol-headw}
\end{figure*}
From \fg{topol-shear} one clearly identifies spiralling streams with the deep atmosphere region, while crossing streams stay at some distance to the planet. For spiralling streams the dominant trend is that they enter from the polar regions ($\cos\theta\approx1$) and leave in the midplane regions. The latter was already seen in the 2D projection of \fg{stream-evol}, but it is with \fg{topol-shear} that the full 3D flow topology becomes clear. The midplane outflow seen in \fg{topol-shear} only forms the base of a `mountain' of outflow trajectories with extends to larger heights. Similarly, for the BS-spiral/in streams there is a noticeable incursion at lower latitudes near $\phi\approx \pi/4$ that almost `touch' the midplane. This valley of incoming and the mountain of exiting spiralling streams are bordered by horseshoe streams in the interior (\ie\ the large white space centred at $\phi=\pi/2$) and non-horseshoe streams on the exterior side (centred at $\phi=0$ and $\phi=\pi$). Thus, the incursions (excursions) into (out of) the deep atmosphere take place near the edge of the horseshoe region. 



The topological map corresponding to the \texttt{headwSph{\hyph}hiRs} run is presented in \fg{topol-headw}. Like in \fg{topol-shear} we see that spiralling streams penetrate most deeply into the atmosphere. However, these streams now enter from the midplane regions while leaving at higher latitudes, reversing the trend of the shear runs.  The spiralling stream in \fg{streams-headw} (purple line with square markers) gives a typical trajectory. First, after entering the BS, it falls in, significantly decreasing its $z$-coordinate. However, during the spiralling it is being lofted to considerable height, where the influence of the `erosive' headwind is stronger and that of the planet's gravity weaker. It is then carried away with the headwind in the $-Y$ direction.

Finally, let us emphasize that \fgs{topol-shear}{topol-headw} are not time-independent. The gray coverage of spiralling trajectories expands and shrinks with time and the fraction of spiralling streams fluctuates strongly. But this variability does not alter the qualitative picture: for the shear-only runs the inflow occurs at high latitudes and the outflow near the midplane and \textit{vice-versa} for the headwind case. 


\begin{figure}
  \includegraphics[width=85mm]{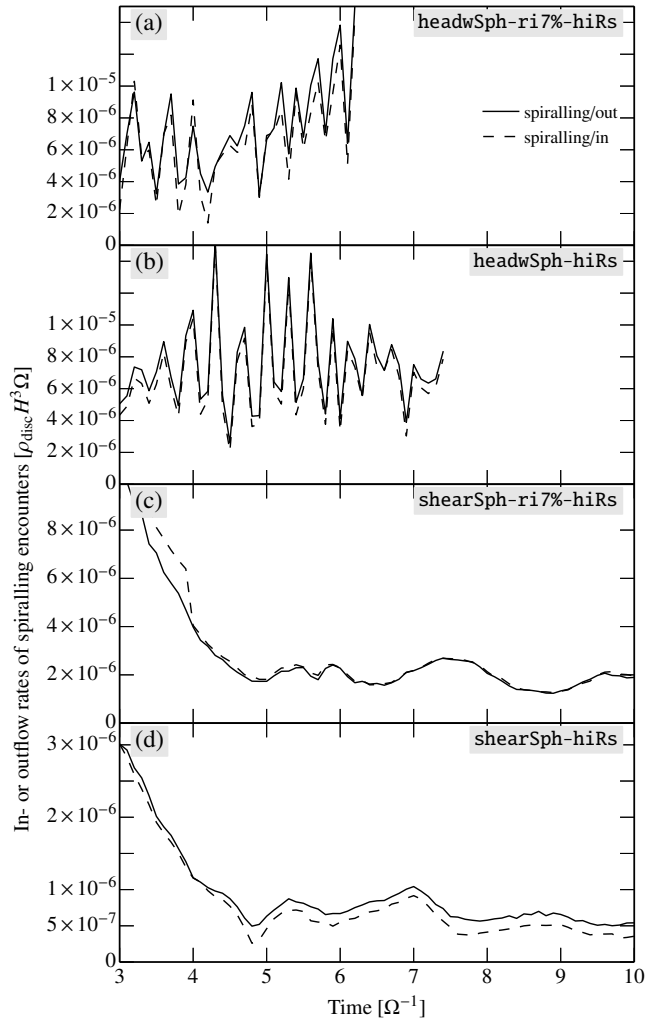}
  \caption{The total rate at which spiralling trajectories -- those that make at least one complete revolution in azimuth -- enter or leave (dashed and solid, respectively) the Bondi sphere. Overall the incoming and outgoing rates match, reflecting mass conservation. Note that the $y$-scaling differs among the panels.}
  \label{fig:integtime}
\end{figure}
\begin{figure*}
  \includegraphics[width=180mm]{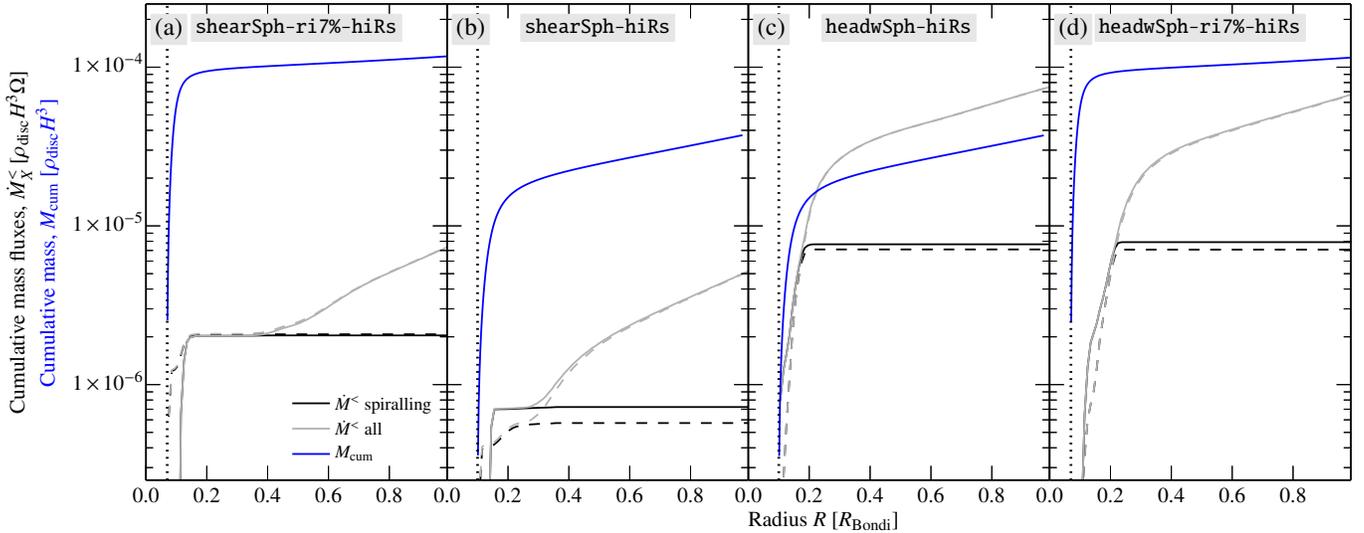}
  \caption{Replenishment rates for the atmosphere. Streams that cross the Bondi sphere are sorted by their minimum distance $r_\mathrm{min}$ to the planet, from which the cumulative mass flux $\dot{M}^<(r)$ of material that reaches radii less than $r$ is calculated. Rates are plotted for the spiralling-only (black) and spiralling+crossing (\ie\ all) streams (gray curves) with solid curves corresponding to the outflow rates and dashed curves to the inflow rates. Values are averaged over times starting from $t=4$. The solid, blue curve gives the cumulative gas mass inside radius $r$. The vertical dashed line gives $r_\mathrm{inn}$.  Comparing $M_\mathrm{cum}(r)$ to $\dot M^<(r)$ one obtains an estimate of the timescale over which the atmosphere gas replenishes.
  }
  \label{fig:replenish}
\end{figure*}
\subsection{The atmosphere replenishment timescale}
\label{sec:replenish}
Next, we calculate the mass flux from streams that enter or leave the BS. For streams that enter the BS (where $v_\mathrm{rad}$ is negative) this reads:
\begin{equation}
  \dot{M}_\mathrm{in} = \frac{4\pi}{N_\theta N_\phi} \sum_{i \in \mathrm{in}} |v_{\mathrm{rad},i}| r_\mathrm{Bondi}^2 \rho_i,
  \label{eq:Mdot-in}
\end{equation}
where the summation can be over a subset (\eg\ only spiralling streams) or over all streams that enter the BS. Similarly, $\dot{M}_\mathrm{out}$ gives the rate at which material leaves the BS. Note that the geometry factor $\sin\theta$ in \eq{Mdot-in} is already accounted for by the equal-area spacing of the grid.

\Fg{integtime} presents these in- and outflow rates corresponding to spiralling streams for all the high resolution runs. Inflow rates are given by a dashed curve while solid curves denotes the amount of gas that leaves the BS. The curves fluctuate strongly in time -- especially those of the headwind runs -- a clear indication that the flow is unsteady. However, the key point of \fg{integtime} is that the in- and outflow rates of spiralling streams match: a jump in $\dot{M}_\mathrm{in}$ is accompanied by a proportional jump of $\dot{M}_\mathrm{out}$. Averaged over time the rates evolve little, except for the headwind \texttt{headwSph-ri7\%-hiRs} run (\cf\ \fg{timeQ}).


\Fg{replenish} presents a different perspective on the outflow rates. Here we have computed the minimum radius $r_\mathrm{min}$ for each of the $N_\theta\times N_\phi$ integrations, sorted these by $r_\mathrm{min}$, and took the cumulative sum such that $\dot{M}^<(r)$ gives the total rate of streams that have reached a minimum distance from the planet no larger than $r$. We have conducted these calculations separately for the three classes of encounters and \textit{time-averaged the rates} for times $t>4\Omega^{-1}$. As the sampling interval is 0.1$\Omega^{-1}$ this implies that several tens of streams are averaged.  In \fg{replenish} we have stacked the $\{\dot{M}^<(r)\}$ for the three classes, so that it shows the in- and outflow rates of spiralling streams (black) and all (light gray) BS-penetrating streams. Dashed lines again denote the inflow rates, whereas solid curves give the outflow rates (in many cases they overlap).

Overall \fg{replenish} confirms the point made above that outflow and inflow rates are (approximately) in balance. We focus here on the radial distribution; \ie\ how deep the streams penetrate into the atmosphere. Clearly one notes that the deeper layers are only probed\? by spiralling streams: $R\lesssim0.4R_\mathrm{Bondi}$ for the shear runs and $R\lesssim0.2R_\mathrm{Bondi}$ for the headwind runs.

Finally, the blue curve in \fg{replenish} gives the cumulative gas mass $M_\mathrm{cum}(R)$ of the atmosphere. The ratio between $M_\mathrm{cum}$ and $\dot{M}$ thus gives an indication of the timescale on which the atmosphere at the particular radius $R$ is being replenished. In the \texttt{headwSph-hiRs} case, for example, the lines cross at $R=0.2R_\mathrm{Bondi}$, which implies that the gas at this location is replenished on a timescale of $\sim$$\Omega^{-1}$. Above this radius replenishment occurs faster. Below slower, but not by much: the gas in the \texttt{headwSph-hiRs} run is only temporarily bound.

This is a general result, although the timescales of the replenishment vary. In the \texttt{shearSph-hiRs} run the accretion rates ($\dot{M}^{<r}$) are lower and it takes longer, up to $10\Omega^{-1}$ to replenish the atmosphere. This explains, why integrations much longer than the potential injection timescale ($t_\mathrm{inj}$ is only 0.5) are required. For the \texttt{shearSph-ri7\%-hiRs} run $t_\mathrm{replenish}$ even reaches $100\Omega^{-1}$ for $r\lesssim0.4$, indicating perhaps that some evolution past $t=10\ \Omega^{-1}$ can be expected.

A striking feature of \fg{replenish} is that the (cumulative) \textit{rates} hardly change when decreasing the inner radius, whereas the cumulative mass $M_\mathrm{cum}$ does. This implies that the replenishment is, to first order, a surface effect and that the efficiency of the replenishment decreases if more mass resides near the interior as is the case for isothermal atmosphere for low $r_\mathrm{inn}$.


We can, crudely, explain the numerical values of the accretion rates from simple geometrical considerations: $\dot{M}(r_\mathrm{min})\sim R_\mathrm{Bondi}^2 v_\mathrm{Bondi} \rho_\mathrm{disc} f_\mathrm{cover}$ where $v_\mathrm{Bondi}$ is the velocity at $R_\mathrm{Bondi}$, and $f_\mathrm{cover}(r_\mathrm{min})$ the fraction of streams that reach a certain minimum depth. For the classical Bondi accretion rate \citep{BondiHoyle1944}, $v_\mathrm{Bondi}=c_s$ and $f_\mathrm{cover}\simeq1$. However, in our case gas is not in free-fall but the pressure balance with the disc ensures that $v_\mathrm{Bondi}$ is given by either the Keplerian shear ($\sim$$\Omega R_\mathrm{Bondi}$) or by the headwind (${\cal M}_\mathrm{hw}c_s$). This leads to a rate $\sim$$r_\mathrm{Bondi}^3\Omega \sim 10^{-6}$ for the shear-only case and ${\cal M}_\mathrm{hw}R_\mathrm{Bondi}^2 c_s\sim 10^{-5}$ for the headwind case. These numbers, which do no account for $f_\mathrm{cover}$, are generally consistent with \fgs{integtime}{replenish}. 

\section{Discussion}
\label{sec:discuss}
\subsection{Implications of the replenishment for atmosphere cooling}
Concluding the above analysis, we have seen that embedded atmospheres of low mass planets represent an open system, where gas continuously enters and leaves the Bondi sphere at rates $\dot{M}$ given by the size of the Bondi sphere, the density of the nebular gas, and the typical velocity at $R_\mathrm{Bondi}$. The corresponding recycling timescale for an atmosphere of mass $M_\mathrm{atm}\equiv\chi_\mathrm{atm} M_p$ is, in the case that the shear dominates the velocities:
\begin{align}
    \nonumber
    t_\mathrm{replenish}
    &\sim \frac{\chi_\mathrm{atm} M_\mathrm{p}}{f_\mathrm{cover}^\ast R_\mathrm{Bondi}^2(R_\mathrm{Bondi}\Omega)\rho_\mathrm{disc}}
    \sim \frac{\chi_\mathrm{atm} h^6 Q_\mathrm{disc}}{f_\mathrm{cover}^\ast q_p^2 \Omega} \\
&\sim 10^4\ \mathrm{yr}\  \frac{\chi_\mathrm{atm}}{f_\mathrm{cover}^\ast} \left( \frac{M_p}{M_\oplus} \right)^{-2} \left( \frac{a}{1\ \mathrm{AU}} \right)^{2.75}.
    \label{eq:replenish}
\end{align}
In the first line, we inserted the definition of the Bondi radius for $R_\mathrm{Bondi}$, with $h$ the aspect ratio of the circumstellar disc, $Q_\mathrm{disc}$ its Toomre's Q parameter, and $q_p = M_\mathrm{p}/M_\mathrm{star}$. In the second line we inserted a standard MMSN disc model \citep{Weidenschilling1977i,HayashiEtal1985}, for which $h^6Q_\mathrm{disc}/\Omega \propto a^{2.75}$. 

In the previous section, we argued that $f_\mathrm{cover}$ depends on the radius, such that $f_\mathrm{cover}(r_\mathrm{min})$ gives the fraction of streams that reach depths $r<r_\mathrm{min}$. Let us for simplicity work with a characteristic fraction $f^\ast_\mathrm{cover}$, which gives $f_\mathrm{cover}$ for $r=r^\ast$ where $r^\ast$ corresponds to the shell where most of the atmospheric mass resides, that is, where $\rho(r) r^3$ peaks. The value of $f^\ast_\mathrm{cover}$ thus depends on the density structure of the atmosphere. For convectively-supported atmospheres $r^\ast$ will lie close to $R_\mathrm{Bondi}$ and $f^\ast_\mathrm{cover}\sim1$. For atmospheres with an (outer) isothermal layer $r^\ast$ and $f_\mathrm{cover}^\ast$ will be lower.


\Eq{replenish} shows, foremost, that atmosphere recycling timescales can become very short -- much shorter than the typical disk lifetimes -- even if $f_\mathrm{cover}^\ast$ is low and $\chi_\mathrm{atm}\approx1$. Therefore, atmospheres of embedded planets should not be regarded as isolated; replenishment will affect the thermal balance of such protoplanets.  Second, \eq{replenish} shows that $t_\mathrm{replenish}$ sensitively depends on the position of the planet in the disc: replenishment is much slower in the outer disk.


The thermal structure of protoplanet atmospheres' is often calculated assuming a certain solid (planetesimals) accretion rate. For a given planet mass, planetesimal accretion rate and atmosphere properties (\eg\ opacity) $\chi_\mathrm{atm}$ will then attain a quasi-static value; and it is only the increase in core mass that drives the evolution \citep[\eg][]{Rafikov2006}. However, if this planetesimal bombardment is insufficient, or shuts off altogether, the luminosity will instead be due to the Kelvin-Helmholtz (KH) contraction of the atmosphere -- a mechanism that increases $\chi_\mathrm{atm}$ with time for constant core mass \citep[\eg][]{IkomaEtal2000}.

One can (numerically) calculate the KH-contraction, or cooling timescale $t_\mathrm{cool}$, which depend on grain and gas properties (opacity and molecular weight), planet mass, and nebular parameters \citep{HoriIkoma2010,BodenheimerLissauer2014,PisoYoudin2014,LeeEtal2014}. For a grain-free atmosphere it can be parameterized as\footnote{\Eq{cool} is an empirical fit, based on the calculations by \citet{HoriIkoma2010} for a grain-free (but not metal-free) atmosphere. The quadratic dependence on $\chi_\mathrm{atm}$ follows from the analytical consideration by \citet{PisoYoudin2014}, although it appears to be steeper in the numerical calculations. Other factors that affect $t_\mathrm{cool}$ are the metalicity of the atmosphere and boundary conditions \citep{LeeEtal2014}. These uncertainties do not affect the argument made here.}
\begin{equation}
    \label{eq:cool}
    t_\mathrm{cool} \sim
    10^7\ \mathrm{yr}\
    \chi_\mathrm{atm}^2 \left( \frac{M_p}{M_\oplus} \right)^{-4.5}.
\end{equation}
When $t_\mathrm{cool}\ll t_\mathrm{replenish}$ atmospheric recycling is slow and does not interfere with the contraction of the atmosphere, allowing $\chi_\mathrm{atm}$ to increase. However, for $t_\mathrm{cool} \gg t_\mathrm{replenish}$ radiative cooling becomes irrelevant, because the atmospheric gas is removed before it cools. The KH-contraction thus ceases when $t_\mathrm{cool}=t_\mathrm{replenish}$ or at an atmosphere mass fraction of:
\begin{equation}
    \label{eq:chi}
    \chi_\mathrm{atm} \sim
    \frac{10^{-3}}{f_\mathrm{cover}^\ast} \left( \frac{M_p}{M_\oplus} \right)^{2.5} \left( \frac{a}{\mathrm{AU}} \right)^{2.75}.
\end{equation}
This equation suggests that grain-free atmospheres of low mass planets ($M_p \lesssim M_\oplus$) do not reach critical values ($\chi_\mathrm{atm}\approx1$) unless $f_\mathrm{cover}^\ast$ turns out to be very low; for the inner disk regions ($\sim$0.1 AU) even planets up to 10 Earth masses can avoid massive atmospheres.

A caveat in the above argument concerns the assumption that the atmosphere can be described by a single value of $f_\mathrm{cover}$. If cooling is effective, as it is initially, and the atmosphere contracts, the deeper dense layers may be reached with more difficulty, as we saw in the $r_\mathrm{inn}=0.07m$ simulation runs. This would reduce $f^\ast_\mathrm{cover}$. However, we never saw the emergence of a region that was truly isolated. In contrast, \citet{LissauerEtal2009}, for parameters tuned to study the formation of Jupiter, found that gas within $R_\mathrm{Hill}/4$ stays bound. This is at odds with our findings; but we note that the \citet{LissauerEtal2009} setup and numerical parameters are very different from ours. Also, their cores are already massive ($\approx$10 Earth masses), and may no longer be fully embedded in the disk.  


None the less, the replenishment is an attractive idea to explain the preponderance of super-Earths and mini-Neptunes: planets up to $\sim$$10\ M_\oplus$ inferred by the \textit{Kepler} mission \citep{FressinEtal2013,PetiguraEtal2013}. Radial-velocity follow up analysis indicates that the densities are relatively low \citep[\eg][]{WeissMarcy2014}. Perhaps the preferred interpretation is that they started with a mass-dominant solid core overlaid by a H/He-rich envelope \citep{WuLithwick2013,OwenWu2013,LopezFortney2014}.  Thus, during the gas-rich circumstellar disc phase these planets somehow avoided crossing the critical core mass, \ie\ their $\chi_\mathrm{atm}$ remained below unity. However, as has been pointed out, accretion of material (planetesimals) in the inner disc regions is rapid \citep{ChiangLaughlin2013}; and when these atmospheres are no longer bombarded by dust-ablating planetesimals, the grain opacity must be negligible \citep{Mordasini2014,Ormel2014}. Consequently, $t_\mathrm{cool}$ is shorter than the disk lifetime for all $\chi_\mathrm{atm}\lesssim1$, implying that these atmospheres should have collapsed to end up as hot-Jupiters. Ideas to prevent the premature collapse of the atmosphere include a short disc lifetime \citep{IkomaHori2012} or a super-solar metallicity atmosphere \citep{LeeEtal2014}, which render $t_\mathrm{cool}(\chi_\mathrm{atm}\sim1)>t_\mathrm{disc}$. Replenishment offers an alternative solution where short cooling timescales are simply irrelevant because of an even more rapid recycling of atmospheric gas. In addition, \eqs{replenish}{chi} show that replenishment is especially important in the inner disc regions -- planets have a hard time to bind gas because they encounter so much of it.

\subsection{Comparison with previous works}

In Paper I we conducted 2D simulations and found circulating streamlines, meaning that the atmosphere is bound. Furthermore, the simulations converged to a steady flow. In this work, by changing the dimensionality of the problem, we no longer find bound material. In addition, we find that the flow pattern is variable. Although the magnitude of the fluctuations are small -- the density obeys the hydrostatic solution -- these in effect break the isolation of the atmospheres. Finally, we find that in 3D, the role of rotation is suppressed compared to the 2D case. 

From this, we must conclude that the dimensionality of the disc greatly influences the results \citep[\cf][]{D'AngeloEtal2002,D'AngeloEtal2003,PaardekooperMellema2006i,PaardekooperMellema2008}. Our results for low mass, embedded protoplanets are generally in agreement with recent literature findings, although these are often more sophisticated regarding their thermodynamic treatment. Like the 3D calculation of this work, \citet{AyliffeBate2009} and \citet{D'AngeloBodenheimer2013} do not observe the formation of disc-like structures. \citet{D'AngeloBodenheimer2013} were, like us, interested in the question where the interface between the material orbiting the planet (bound to its atmosphere) and orbiting the star lies. 
They find that this interface lies at $\approx$$0.4R_\mathrm{Bondi}$ for their 10 AU runs. Interestingly, this value roughly corresponds to the point where spiralling streams occur in our calculations (see \fg{replenish}).  However, a difference is that we do observe material from deeper layers to eventually escape the atmosphere. It is not clear what the origin of this difference is. Most likely, it is due to different physics the respective simulations include; \eg\ where we use an isothermal EOS and an inviscid gas, \citet{D'AngeloBodenheimer2013} solve for the radiation transport and use explicit viscosity. In addition, they use a global setup with a nested grid that refines towards the planets, instead of the local linearized geometry of this paper. In a subsequent work, we will investigate whether these conclusions alter when adopting a more realistic treatment of the thermal balance, instead of simply prescribing an EOS.

Different from these works, and more in line with this study, \citet{WangEtal2014} find an open system, where material is continuously replenished. The \citet{WangEtal2014} simulations are perhaps most similar to this study as they also adopt an isothermal EOS and omit self-gravity and explicit viscosity\?. They also focus on low-mass planets, although their smallest mass -- 4 Earth masses at 5.2 AU or $m=0.1$ in dimensionless units -- is still a factor 10 larger than the $m=10^{-2}$ adopted here. However, \citet{WangEtal2014} is unique in finding evidence of a circumplanetary disk for these low mass planets, which implies that material is rotating near-Keplerian. It is possible that this transition to Keplerian rotation for inviscid flows may occur for these $m\gtrsim0.1$ planets or for more massive atmospheres (\ie\ if we adopt an even lower inner radius) as we saw in Paper I. However, we do expect that for low-mass planets ($m<1$) such a transition would occur smoothly and does not involve non-steady spiral shocks as seen by \citet{WangEtal2014}.

To address these issues it is important to conduct simulations for higher mass planets. However, for planets of mass $m\gtrsim0.1$ the adopted local and spherical geometry becomes problematic. The premise of these local simulations is that the conditions at the outer boundary, \ie\ at sufficiently large distance from the planet, are given by the unperturbed flow. Phenomena as gap opening or flattening (circumplanetary disk formation) cannot be followed. For gap opening, which becomes important for high-mass $m\gtrsim1$ planets, one must adopt a global setup that follows the full coorbital region. However, the vertical settling of gas near the planet can still be followed locally in Cartesian or cylindrical geometries by adopting a large enough domain in $z$ and changing the boundary condition of the $z$-axis to reflective, thereby conserving the surface density. But this option is not available for the spherical grid. For these reasons we have limited ourselves to the well-embedded, $m\ll1$, case.

\section{Summary}
\label{sec:summary}
In this paper, we have conducted hydrodynamical simulations to study the flow pattern past small, embedded planets by employing a spherical grid centred on the planet and characterized by an inner boundary -- a setup that was very beneficial to study 2D flows. 
However, the 3D calculations have turned out to be considerably more challenging than their 2D counterparts. Numerically, the spherical grid's inhomogeneous spacing forces a small timestep. In addition, as the 3D flow no longer conserves vortensity the fidelity of the simulation results are more difficult to ascertain. Anoter formidable drawback is that the simulations take much longer to converge to their long-term solution. The underlying reason for the long timescales is that material takes a long time (hydrodynamically speaking) to cross the atmosphere.

We summarize the key findings:
\begin{enumerate}
    \item Compared to the 2D case, atmospheres tend to be less dominated by rotation. The vertical (compressive) motions are effective in quenching the vortensity (\fg{vortensity}). Nevertheless, we do find that the rotation increases for more massive atmospheres, when the inner radius of the simulations -- a proxy for the density of the atmosphere -- is reduced, see \fg{rhophi}. This hints at a possible transition to a more disc-like configuration for higher mass atmospheres.
    \item The velocity field is time-variable -- with the fluctuations being a substantial fraction of the mean flow velocity. However, but for the low-mass planets studied here the fluctuations fall in the subsonic regime. As a result, the atmosphere continuously exchanges mass with the circumstellar disc. 
    \item We have characterized streamlines that enter the Bondi sphere (BS) into crossing and spiralling (see \fgs{streams-headw}{streams-shear}). Spiralling streamlines make one or more revolution within the BS and penetrate most deeply into the atmosphere.
    \item In particular, in the shear-only setup (where the gas on the planet's orbit corotates with the planet) material enters the Bondi sphere at mid-to-high latitudes, and exits the Bondi sphere in the midplane regions. For the headwind runs (where the gas is pressure supported to move at sub-Keplerian velocities) a reverse trend is seen: material mostly enters in the midplane regions, and leaves at higher latitudes (\fgs{topol-headw}{topol-shear}).
    \item Generally, the atmosphere replenishment rates can be understood by geometrical arguments, which leads to a modified Bondi rate of $\dot{M} \sim R_\mathrm{Bondi}^3 \Omega \rho_\mathrm{disc}$ for the shear-only case and $\dot{M}\sim R_\mathrm{Bondi}^2 {\cal M}_\mathrm{hw} c_s\rho_\mathrm{disc}$ for the headwind case. For the deeper layers these rates should be multiplied by a coverage fraction $f_\mathrm{cover}(r_\mathrm{min})$, which characterizes the fraction of streams that reach a certain depth $r_\mathrm{min}$. Here, we find that $f_\mathrm{cover}\sim0.1$--1.
    \item These estimates result in an atmosphere replenishment timescales $t_\mathrm{replenish}$ which are short compared to the cooling timescale of the atmosphere, especially for planets in the inner parts of the disks where $\rho_\mathrm{disc}$ and $\Omega^{-1}$ are large. As a result, the contraction of the atmospheres of such protoplanets stalls at an atmosphere mass fraction $M_\mathrm{atm}/M_p\ll1$.
\end{enumerate}

The last point conclusion is highly timely as it offers a possible explanation for the preponderance of super-Earths and mini-Neptunes -- systems that are presumably gas-rich but rock-dominant. We admit that this constitutes a leap of faith, as the simulations of this work pertain a quite different regime (very low-mass planets) and neglect realistic thermodynamics. We therefore encourage more studies in this field. At the very least, this work has demonstrated that atmospheres of embedded planets are not necessarily isolated, as has been historically the standard view \citep[\eg][]{Mizuno1980,PollackEtal1996}. If this finding proves robust, it will change the understanding of the early evolution of low mass planets.





\section*{Acknowledgments}
This work has profited immensely from discussion with many colleagues. The simulations were performed with the Berkeley cluster \texttt{Henyey}, which was made possible by a National Science Foundation Major Research Instrumentation (NSF MRI) grant. For CWO support for this work was provided by NASA through Hubble Fellowship grant \#HST-HF-51294.01-A awarded by the Space Telescope Science Institute, which is operated by the Association of Universities for Research in Astronomy, Inc., for NASA, under contract NAS 5-26555.  RK acknowledges funding from the Max Planck Research Group `Star formation throughout the Milky Way Galaxy' at the Max Planck Institute for Astronomy. 

\bibliographystyle{mn2e}
\bibliography{ads,arXiv,mybibl}

\appendix
\section[]{The gravitational force on the inner boundary}
\label{app:innerB}
\begin{figure}
  \includegraphics[width=85mm]{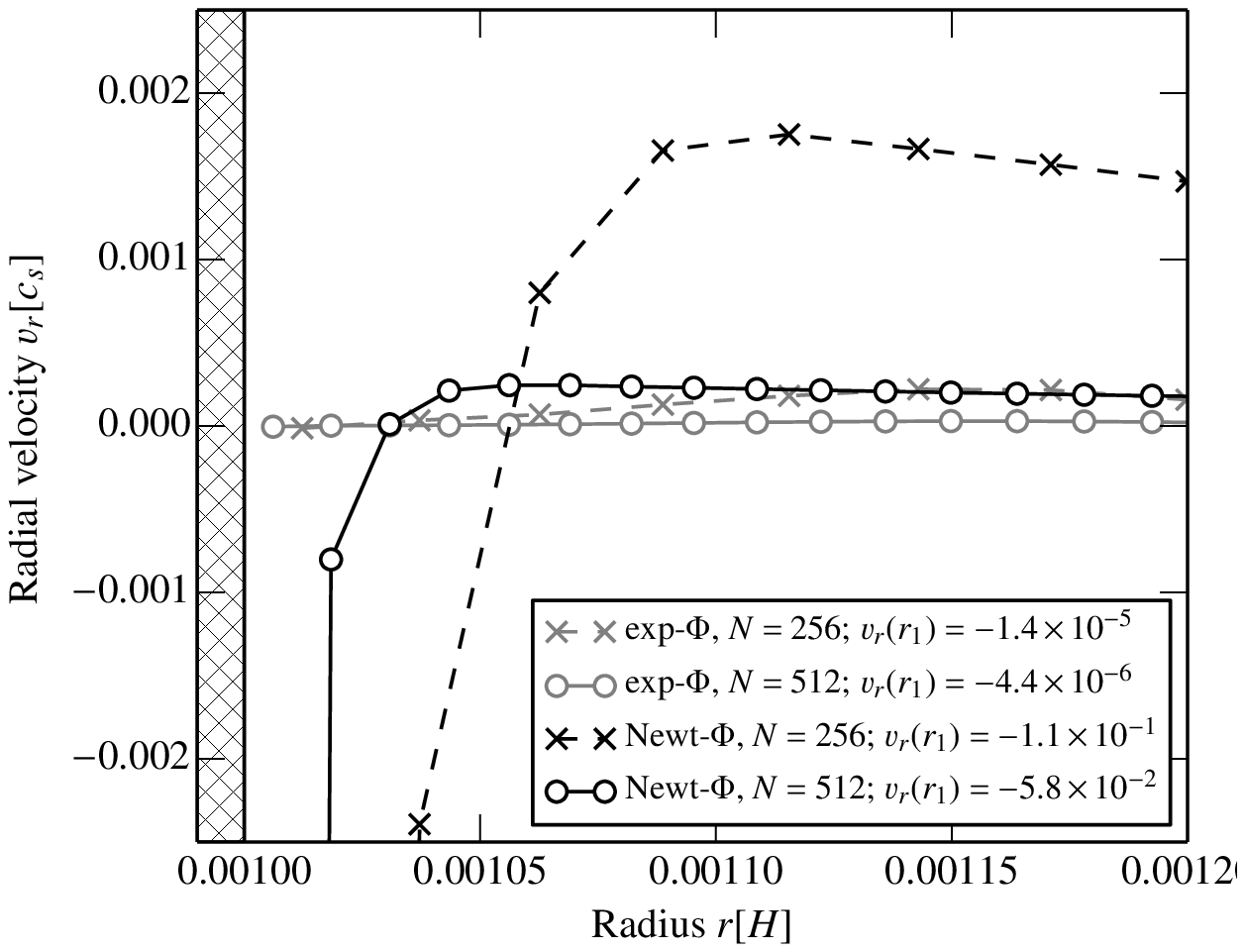}
  \caption{Excess radial motions after $t=10$ for a test calculations where gas settles according to a prescribed gravitational potential. Grid spacing and units are the same as in the main text, but the potential is only 1D and includes a damping term proportional to velocity. Black curves: Newtonian potential; blue curves: $\Phi_\mathrm{2b}$ (\eq{Phi-2b}). Symbols denote different grid resolution. For the Newtonian potential spurious velocities emerge near the inner boundary.}
  \label{fig:collapse}
\end{figure}
We motivate the choice of a force-free inner boundary, \eq{F2b}, by showing that the standard Newtonian potential in combination with a reflective boundary results in artificially strong radial motions close to the boundary. Here, the setup is a simplified 1D version of the equations described in \se{model} but tailored to isolate this particular problem. Non-inertial forces are omitted; only the two-body force is used, which we inject slowly as in \eq{F2b-time}. The grid is, as in the main text, exponentially spaced and extends from $r_\mathrm{inn}=10^{-3}$ to $r_\mathrm{out}=0.5$. In addition, we include a damping force, $F_\mathrm{damp}=-v$, with the aim of damping any excess motion. Therefore, after the infall phase has completed, the density is expected to relax to the hydrostatic solution, $\rho=\exp[-\Phi_\mathrm{2b}]$, and the velocities should vanish everywhere, $v=0$.

The results for two resolutions, 256 and 512 radial grid points, and for the two potentials, Newtonian ($\Phi_\mathrm{Newt}=-m/r$) and the exponential potential of \Eq{Phi-2b} (see \fg{potentials}), are shown in \fg{collapse}. The figure plots the excess radial velocity after a time $t=10$. For the exponential potential the velocities all lie close to zero, as anticipated. The higher resolution run converges better to zero than the lower resolution.

However, for the Newtonian potential velocities deviate significantly from zero, especially near the inner boundary, where the radial motions is $v\approx-0.11 c_s$ (see the legend for the value of the velocity at the first grid point). It seems, then, that there is a flow out of the domain, which is at odds with the reflective boundary conditions. On the other hand, \pluto\ does retrieve a steady state: density and velocities have become time-independent. 


These findings are obviously boundary effects: after the 2nd grid point the solution converges more quickly. In addition, a higher resolution improves the discrepancy. However, for the multi-dimensional problem of the main paper these boundary effects feed back onto the main flow, rendering it unsteady. Clearly, to remedy these concerns, the boundary condition must be modified. It is unclear, however, which boundary condition yields the best representation of the static isothermal atmosphere solution. Numerically, the easiest approach seems to be the modification of the Newtonian force at the domain interface, similar to a softening of the potential often used in $N$-body and SPH simulations. This is achieved by adopting the exponential, near-Newtonian potential $\Phi_\mathrm{2b}$ of \eq{Phi-2b}.


\bsp
\label{lastpage}
\end{document}